\documentclass[pdflatex,sn-nature]{sn-jnl}

\usepackage{graphicx}%
\usepackage{multirow}%
\usepackage{amsmath,amssymb,amsfonts}%
\usepackage{amsthm}%
\usepackage{mathrsfs}%
\usepackage[title]{appendix}%
\usepackage{xcolor}%
\usepackage{textcomp}%
\usepackage{manyfoot}%
\usepackage{booktabs}%
\usepackage{lineno}
\usepackage{algorithm}%
\usepackage{algorithmicx}%
\usepackage{algpseudocode}%
\usepackage{listings}%

\usepackage{amssymb}
\usepackage{makecell} 
\usepackage{rotating} 

\theoremstyle{thmstyleone}%
\theoremstyle{thmstyletwo}%

\theoremstyle{thmstylethree}%

\begin{document}

\title[Article Title]{\texttt{DiffSyn:} A Generative Diffusion Approach to Materials Synthesis Planning}


\author[1]{\fnm{Elton} \sur{Pan}}\email{eltonpan@mit.edu}

\author[2]{\fnm{Soonhyoung} \sur{Kwon}}\email{s1105hk@mit.edu}

\author[1]{\fnm{Sulin} \sur{Liu}}\email{sulinliu@mit.edu}

\author[2]{\fnm{Mingrou} \sur{Xie}}\email{mrx@mit.edu}

\author[1]{\fnm{Alexander  J.} \sur{Hoffman}}\email{ajhoff29@mit.edu}

\author[1]{\fnm{Yifei} \sur{Duan}}\email{duanyf99@mit.edu}

\author[3]{\fnm{Thorben} \sur{Prein}}\email{t.prein@tum.de}

\author[1]{\fnm{Killian} \sur{Sheriff}}\email{ksheriff@mit.edu}

\author[2]{\fnm{Yuriy} \sur{Roman-Leshkov}}\email{yroman@mit.edu}

\author[4]{\fnm{Manuel} \sur{Moliner}}\email{mmoliner@itq.upv.es}

\author[1]{\fnm{Rafael} \sur{Gomez-Bombarelli}}\email{rafagb@mit.edu}

\author*[1]{\fnm{Elsa} \sur{Olivetti}}\email{elsao@mit.edu}

\affil[1]{\orgdiv{Department of Materials Science and Engineering}, \orgname{Massachusetts Institute of Technology}, \orgaddress{\city{Cambridge}, \state{MA}, \postcode{02139}, \country{USA}}}

\affil[2]{\orgdiv{Department of Chemical Engineering}, \orgname{Massachusetts Institute of Technology}, \orgaddress{\city{Cambridge}, \state{MA}, \postcode{02139}, \country{USA}}}

\affil[3]{\orgdiv{Department of Chemistry}, \orgname{Technische Universität München}, \orgaddress{\city{München}, \postcode{80333}, \country{Germany}}}

\affil[4]{\orgname{Instituto de Tecnolog\'{i}a Qu\'{i}mica, Universitat Polit\`{e}cnica de Val\`{e}ncia-Consejo Superior de Investigaciones Cient\'{i}ficas}, \orgaddress{\city{Valencia}, \postcode{46022}, \country{Spain}}}


\abstract{The synthesis of crystalline materials, such as zeolites, remains a significant challenge due to a high-dimensional synthesis space, intricate structure-synthesis relationships and time-consuming experiments. Considering the \textit{one-to-many} relationship between structure and synthesis, we propose \texttt{DiffSyn}, a generative diffusion model trained on over 23,000 synthesis recipes spanning 50 years of literature. \texttt{DiffSyn} generates probable synthesis routes conditioned on a desired zeolite structure and an organic template. \texttt{DiffSyn} achieves state-of-the-art performance by capturing the multi-modal nature of structure-synthesis relationships. We apply \texttt{DiffSyn} to differentiate among competing phases and generate optimal synthesis routes. As a proof of concept, we synthesize a UFI material using \texttt{DiffSyn}-generated synthesis routes. These routes, rationalized by density functional theory binding energies, resulted in the successful synthesis of a UFI material with a high  Si/Al$_{\text{ICP}}$ of 19.0, which is expected to improve thermal stability and is higher than that of any previously recorded.
}





\maketitle

\section{Introduction}\label{sec1}

Materials discovery lays the foundation for modern technologies, from catalysis to electronics \cite{butler2018machine}. Recent large-scale computational searches of chemical composition and structures \cite{merchant2023scaling, sriram2024open, zhu2024uncovering, kim2021solid} have uncovered millions of potentially stable, synthesizable materials (\textit{what} to synthesize) \cite{merchant2023scaling, zeni2023mattergen, barroso2024open, saal2013materials}. However, finding viable synthesis routes remains a bottleneck in materials discovery (\textit{how} to synthesize) \cite{he2023precursor, huo2022machine, karpovich2023interpretable, mahbub2020text} because there are many synthesis parameters (composition, conditions, etc.) that interact in complex ways. Moreover, the compute required for atomistic simulations scales poorly with system size, precluding accurate modeling of the underlying physical phenomena in complex materials (i.e., thermodynamics and kinetics) \cite{bianchini2020interplay}. Consequently, there is continued interest in machine learning (ML) approaches to directly learn from experimental synthesis data to predict materials synthesis parameters at lower computational cost \cite{huo2022machine, karpovich2023interpretable, pan2024zeosyn}.

Materials synthesis prediction presents a unique challenge for ML for several reasons. First, structure-synthesis relationships are \textit{one-to-many}---i.e., a single target structure may form through \textit{multiple} possible synthesis recipes. Second, the inverse relationship (synthesis-structure) is also one-to-many---i.e., a single recipe may result in the formation of a mixture of products (competing phases) due to the complex interplay of thermodynamic and kinetic pathways \cite{bianchini2020interplay}. Capturing this phase competition is crucial to selectively synthesize single-phase materials instead of mixtures. Third, complex non-linear interactions exist between synthesis parameters, such as temperature and time (Fig. \ref{fig:physical_relationship}) \cite{karpovich2023interpretable}, requiring approaches that model joint probabilities across multiple synthesis parameters. Predictions must capture relationships among variables to make tradeoffs between parameters leveraging physical information about materials synthesis (e.g., crystallization kinetics).




Previous ML approaches to predict synthesis have predominantly used regression approaches \cite{karpovich2023interpretable, prein2024reaction, schwalbe2023inorganic, luo2022mof}, which deterministically map a representation (e.g., composition \cite{huo2022machine, karpovich2023interpretable}, structural features \cite{schwalbe2023inorganic}, graphs \cite{prein2024reaction}) of a material to its synthesis parameters. These approaches are limited because the deterministic mapping is incompatible with the one-to-many nature of structure-synthesis relationships and assume independence between synthesis parameters (Fig. \ref{fig:2d_dist}) \cite{pan2024chemically, karpovich2021inorganic}. 
These factors limit the predictive accuracy of regression approaches and motivate a shift to generative models, which can sample a complex distribution that accounts for the non-linear interactions between parameters in high-dimensional synthesis space.

To address these challenges, we introduce a diffusion model for materials synthesis. Diffusion models are a powerful class of generative models that were demonstrated to generate novel, high-quality images conditioned on text \cite{saharia2022photorealistic, ramesh2021zero}. Diffusion models can be guided at each step of the denoising process toward a specific objective (e.g., target material) \cite{ho2022classifier, dhariwal2021diffusion}. Unlike generative adversarial networks that suffer from mode collapse \cite{zhang2018convergence}, diffusion models can generate diverse outputs because they are trained to denoise data (e.g., a synthesis route) that have been corrupted with noise. Other generative approaches, namely variational autoencoders and normalizing flows, have limited expressivity due to one-step decoding and affine invertible layers, respectively. In contrast, the iterative denoising process renders diffusion models highly expressive, which enables high sample quality \cite{dhariwal2021diffusion}. This high expressivity may enable diffusion models to capture boundaries in synthesis space between competing phases.

\begin{figure}[t]
    \centering
    \includegraphics[width=\linewidth]{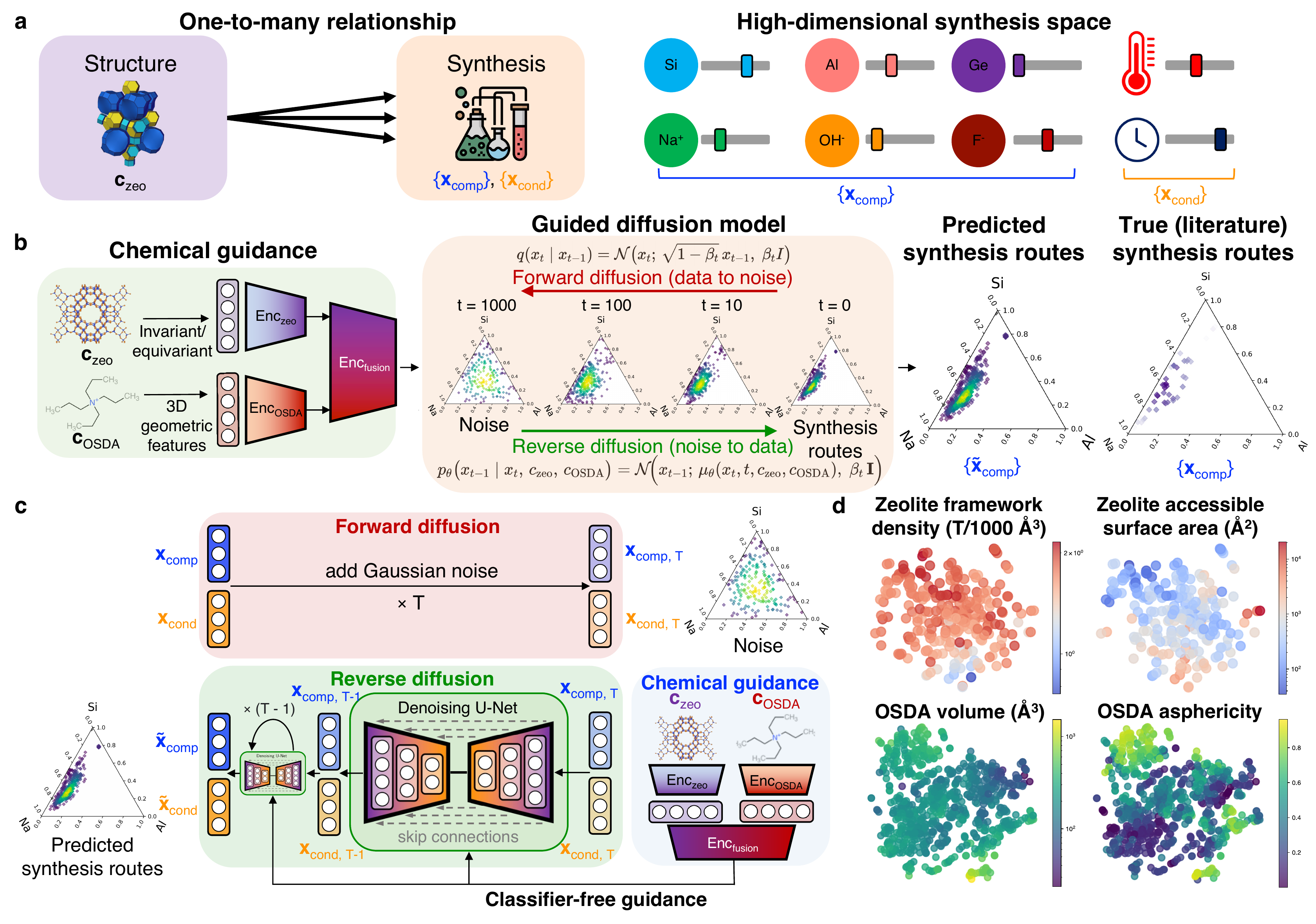}
    \caption{\textbf{\texttt{DiffSyn:} A diffusion approach to materials synthesis planning}. \textbf{(a)} Materials, such as zeolites, often have an \textit{one-to-many} structure-synthesis relationship, where a structure $c_{\text{zeo}}$ can be synthesized via multiple possible synthesis routes in a high-dimensional synthesis space (compositions $\{x_{\text{comp}}\}$ and conditions $\{x_{\text{cond}}\}$).
    \textbf{(b)} Overview of \texttt{DiffSyn}: Given a desired zeolite structure $c_{\text{zeo}}$ and an organic structure directing agent $c_{\text{OSDA}}$, the model $p_{\theta}(x_{\text{comp}}, x_{\text{cond}}|c_{\text{zeo}}, c_{\text{OSDA}})$ generates an \textit{ensemble} of synthesis routes $\{\tilde{x}_{\text{comp}}\}$ and $\{\tilde{x}_{\text{cond}}\}$ via reverse diffusion (green arrow) that matches ground-truth synthesis routes $\{x_{\text{comp}}\}$, accurately capturing the aforementioned one-to-many structure-synthesis relationship with which previous methods struggle. Note: Only $\{\tilde{x}_{\text{comp}}\}$ is shown. \textbf{(c)} Model architecture: Two separate encoders embed the zeolite material $c_{\text{zeo}}$ and organic template $c_{\text{OSDA}}$ before their latent representations are concatenated. The joint representation steers the generation process from noise ($x_{\text{comp}, T}, x_{\text{cond}, T} \sim \mathcal{N}(0, \mathbf{I})$) to realistic synthesis routes specific to the zeolite and OSDA via classifier-free guidance by learning $\mu_\theta\left(x_t, t, c_{\text{zeo}}, c_{\text{OSDA}}\right)=\frac{1}{\sqrt{\alpha_t}}\left(x_t-\frac{\beta_t}{\sqrt{1-\bar{\alpha}_t}} \epsilon_\theta\left(x_t, t, c_{\text{zeo}}, c_{\text{OSDA}}\right)\right)$ where $\epsilon_\theta\left(\cdot\right)$ is a conditional U-Net trained to predict the noise at time $t$ (Section \ref{ddpm}). Note that $x_{\text{comp}}$ and $x_{\text{cond}}$ are jointly noised and denoised. \textbf{(d)} PCA of learned representations of zeolites and OSDAs. The colorbar refers to a specific physical property (as shown in each title).}
    \label{fig:fig1}
\end{figure}

We propose \texttt{DiffSyn}, a diffusion approach to materials synthesis prediction and demonstrate it on zeolites, which are crystalline, microporous materials with applications in catalysis, adsorption, and ion exchange. Zeolite synthesis is challenging due to its high dimensionality (Fig. \ref{fig:fig1}a), with numerous variables influencing the synthesis outcome (Fig. \ref{fig:osda_templating}). Moreover, multiple modes of valid synthesis routes exist for a given structure (Fig. \ref{fig:2d_pca_compare_with_struct}). Progress in zeolite synthesis has focused on trial-and-error experiments guided by domain heuristics \cite{moliner2019machine}. We demonstrate the first application of a guided diffusion model for materials synthesis prediction, which exhibits state-of-the-art performance compared to regression-based and other deep generative approaches. We show that the performance of \texttt{DiffSyn} arises from its ability to capture the one-to-many and multi-modal nature of structure-synthesis relationship in materials. We experimentally validate our approach by synthesizing the UFI zeolite based on \texttt{DiffSyn}-generated synthesis routes. We rationalize these routes using density functional theory (DFT) calculations of inorganic cations that guide UFI synthesis. Together, these results indicate that \texttt{DiffSyn} learns the underlying chemistry that influences synthesis outcomes implicitly from published synthesis recipes.

\section{Results}

\subsection{
\texttt{DiffSyn} framework for materials synthesis planning}

\subsubsection{Chemically guided diffusion model}
\texttt{DiffSyn} leverages a chemically guided diffusion model for predicting materials synthesis routes with the target zeolite structure $c_{\text{zeo}}$ and the organic structure-directing agent (OSDA) $c_{\text{OSDA}}$ as inputs (Fig. \ref{fig:fig1}b). An OSDA is an organic molecule that templates the zeolite's pores (Fig. \ref{fig:osda_templating}e), directing the synthesis toward the formation of a specific structure. Prior work has shown that the optimal OSDA to synthesize a given structure can be identified from atomistic simulations \cite{lewis1995predicting, zones1996guest, jensen2021discovering, schwalbe2021priori, hoffman2024learning, xie2024exhaustive}. The goal is to learn $p(x_{\text{comp}}, x_{\text{cond}}|c_{\text{zeo}}, c_{\text{OSDA}})$ to generate an \textit{ensemble} of synthesis routes consisting of gel compositions $\{x_{\text{comp}}\}$ and synthesis conditions $\{x_{\text{cond}}\}$ given a target structure and OSDA, as shown in Fig. \ref{fig:task} (also known as synthesis parameters, defined in Table \ref{thresholds}). Fig. \ref{fig:2d_dist} shows an example of the predicted vs. ground truth synthesis parameters (green points). The prediction of synthesis parameters is an under-determined problem with multiple possible valid synthesis routes $\{x_{\text{comp}}, x_{\text{cond}}\}$ for each $c_{\text{zeo}}$ and $c_{\text{OSDA}}$.

During training, the forward diffusion process (red arrow in Fig. \ref{fig:fig1}b) adds Gaussian noise to $x_{\text{comp}}$ and $x_{\text{cond}}$, progressively mapping them to a Gaussian distribution (noise). During inference, the reverse diffusion process (green arrow in Fig. \ref{fig:fig1}b) starts from Gaussian noise and iteratively denoises using a U-Net \cite{ronneberger2015u} conditioned on chemical guidance (see Section \ref{representation}) via classifier-free guidance (Fig. \ref{fig:fig1}c). After $T$ timesteps of denoising, the model generates synthesis routes for a desired structure. This denoising process can be seen in the improvement of generation metrics (e.g., Wasserstein distance and COV-P, defined in Section \ref{section:metrics}) throughout the reverse diffusion process (Fig. \ref{fig:metrics_vs_t}). We train \texttt{DiffSyn} on the ZeoSyn dataset \cite{pan2024zeosyn}, which consists of 23,961 synthesis recipes, 233 zeolite topologies, and 921 OSDAs (Supplementary Section \ref{zeosyn_dataset}).

\subsubsection{Representation learning of materials} \label{representation}
\texttt{DiffSyn} integrates a dual-encoder approach consisting of separate encoders (Enc$_{\text{zeo}}$ and Enc$_{\text{OSDA}}$) for the zeolite structure and OSDA, respectively (Fig. \ref{fig:fig1}b). We use two representations of the zeolite structure: invariant geometric features and an equivariant graph neural network (EGNN). The invariant geometric features are physical descriptors (e.g., pore volume) calculated from the zeolite structure using the \texttt{Zeo++} package. The EGNN encoder directly learns a representation from a graph of the zeolite crystal structure (Supplementary Section \ref{model_implementation}). For the OSDA, we perform molecular geometry relaxation and calculate its physicochemical descriptors (e.g., volume and shape) (Section \ref{featurization}). 

Fig. \ref{fig:fig1}d shows that the respective encoders learn smooth and continuous latent spaces with respect to the properties of zeolites and OSDAs. A comprehensive set of properties plotted in embedding space can be found in Fig. \ref{fig:zeolite_embeddings} and \ref{fig:osda_embeddings}, indicating chemically meaningful representations of zeolites and OSDAs. These representations are concatenated before a fusion encoder (Enc$_{\text{fusion}}$) learns a joint representation. We refer to the joint representation as chemical guidance (Fig. \ref{fig:fig1}b). Chemically meaningful representations are pivotal in steering the diffusion model to generate realistic synthesis routes for a desired materials structure. This representation enables \texttt{DiffSyn} to generate synthesis parameters that reflect synthesis routes unseen in training, which have been reported in the literature (Fig. \ref{fig:fig2}e).

\subsubsection{Influence of chemical guidance in diffusion model}

Classifier-free guidance \cite{ho2022classifier} is a critical component of \texttt{DiffSyn}, where the chemical guidance steers the generation process by reweighing the unconditional score function with a conditional score function (Section \ref{sec:cfg}). We probe the influence of two key hyperparameters related to classifier-free guidance: the probability of the chemical guidance being omitted in score estimation during training ($p_\text{uncond}$) and the guidance strength that weighs the conditional score relative to the unconditional score during inference ($w$; Eq. \ref{eq:cfg}).

Higher values of $p_\text{uncond}$ and $w$ amplify the conditional score, making the sampling process more dependent on the chemical guidance. This increased dependence on chemical guidance can result in over-constrained outputs, reducing diversity and potentially missing valid synthesis routes. Lower values of $p_\text{uncond}$ and $w$ reduce the influence of the chemical guidance, potentially decreasing specificity and generating synthesis recipes that are less tailored to the target structure, but improving the diversity of recipes. We find that the balance between diversity and quality of generated synthesis routes occurs at  $p_\text{uncond}$ = 0.1 and $w$ = 1.0 (Fig. \ref{fig:p_uncond_and_cond_scale}).

\subsection{Modeling structure-synthesis relationships}
We evaluate \texttt{DiffSyn} against a suite of previously published approaches in materials synthesis planning. These baseline models fall into three categories: regression-based approaches (average minimum distance (AMD) \cite{schwalbe2023inorganic} and Bayesian neural networks (BNN) \cite{jospin2022hands}); classical generative models (Gaussian mixture model (GMM) \cite{reynolds2009gaussian}); and deep generative models (conditional generative adversarial network (GAN) \cite{mirza2014conditional}, normalizing flow (NF) \cite{dinh2016density}, and variational autoencoder (VAE) \cite{karpovich2023interpretable, karpovich2021inorganic}). More information about these baselines is included in Section \ref{section:baseline}. We compare approaches using Wasserstein distance (lower is better), which measures the distance between the generated and ground-truth distributions of literature reported synthesis parameters for unseen zeolite-OSDA systems. Additionally, we propose the coverage metric COV-F1. The model should maximize both COV-P (precision) and COV-R (recall) simultaneously. Therefore, their harmonic mean (COV-F1) measures the degree of generated recipes being both realistic and diverse (ranges from 0--1, higher is better). Detailed explanations of these metrics are in Section \ref{section:metrics} and Supplementary Section \ref{coverage_metrics}.

\subsubsection{Generative approaches better model structure-synthesis relationships}
\begin{figure}[H]
    \centering
    \includegraphics[width=\linewidth]{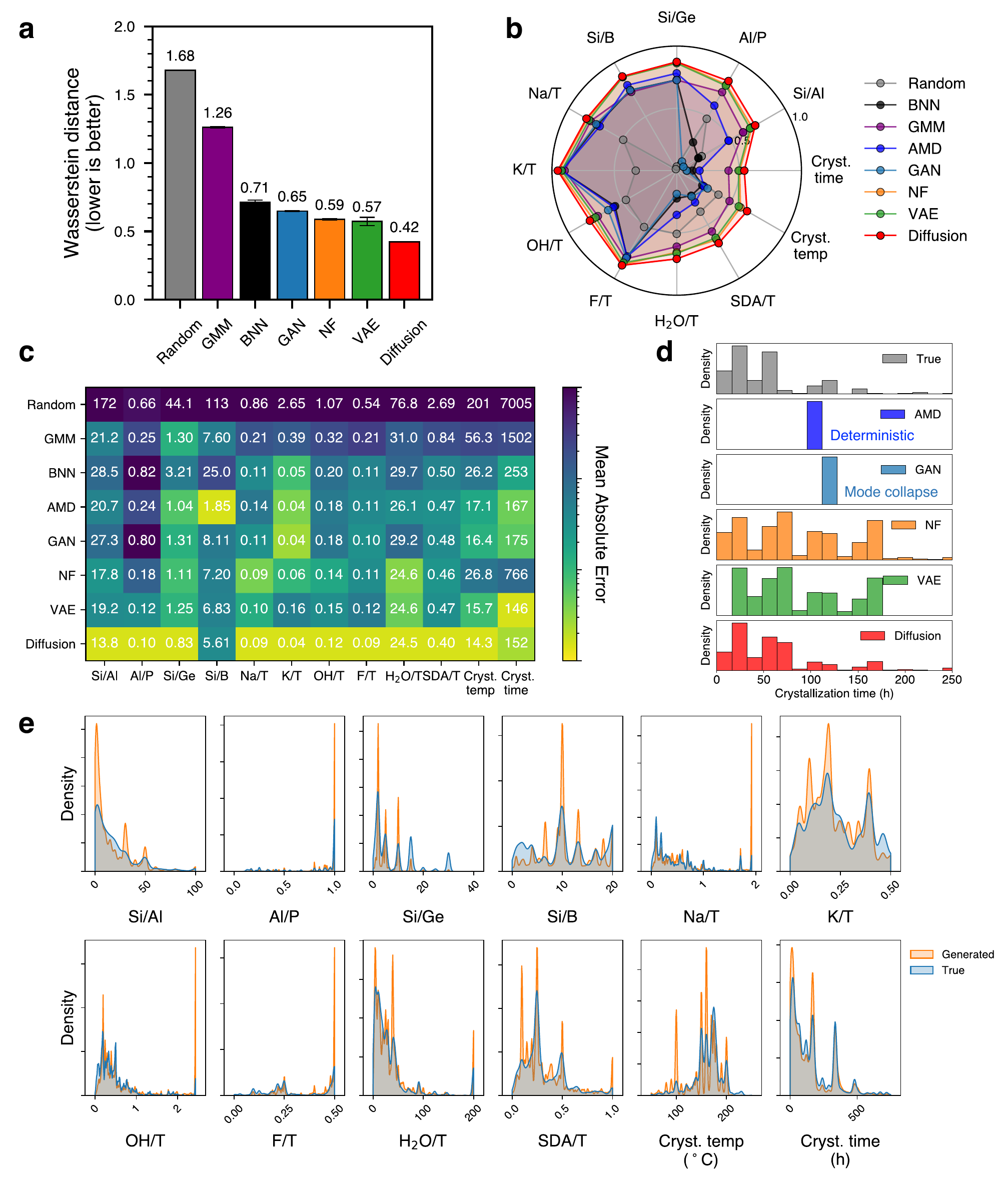}
    \caption{\textbf{$\texttt{DiffSyn}$ achieves state-of-the-art performance in materials synthesis prediction}. \textbf{(a)} Wassterstein distance (lower is better) between generated and literature synthesis routes. Metric is averaged across all test systems. Error bars correspond to standard deviation across 3 independent training runs. \textbf{(b)} COV-F1 (higher is better, ranging from 0 to 1) of individual synthesis parameters. \textbf{(c)} Mean absolute error between the means of distribution of generated and literature synthesis parameters. \textbf{(d)} Distributions of predicted and true synthesis parameters across various different modeling approaches. AMD (purple) is regression-based (outputs deterministic, single-point prediction), while GAN (light blue) suffers from mode collapse. NF and VAE output distributions do not match the ground-truth (grey). $\texttt{DiffSyn}$ (red) accurately captures the true distribution (grey) of the synthesis parameters. \textbf{(e)} $\texttt{DiffSyn}$-generated synthesis routes (orange) and distribution of synthesis routes reported in literature (blue). Synthesis parameters are aggregated across all unseen zeolite-OSDA systems.
    }
    \label{fig:fig2}
\end{figure}

Wasserstein distances show that deep generative models such as GAN, NF, VAE and $\texttt{DiffSyn}$ outperform the classical approaches, with $\texttt{DiffSyn}$ outperforming the next best baseline (VAE) by over 25$\%$. Classical generative approaches like GMM do not perform much better than a random baseline (Fig. \ref{fig:fig2}a) while a probabilistic regression model (BNN) performs better than GMM.

The COV-F1 of 12 synthesis parameters are shown in Fig. \ref{fig:fig2}b. The models perform better on synthesis parameters related to heteroatoms (Si/Al, Al/P, Si/Ge, Si/B), cations (Na$^+$/T, K$^+$/T) and anions (F$^-$/T, OH$^-$/T). However, they struggle to predict crystallization time, which could be attributed to anthropogenic factors; crystallization times are subject to human bias, where experimentalists test and report “rounded” numbers \cite{jia2019anthropogenic}. This bias results in the ground-true time distribution peaking at specific intervals, hence rendering time prediction more challenging. 

Deep generative approaches (VAE, NF, and \texttt{DiffSyn}) outperform regression-based approaches (AMD and BNN). We hypothesize that generative models have superior performance due to better recall (higher COV-R, Fig. \ref{fig:precision_recall}). Meanwhile, $\texttt{DiffSyn}$ outperforms other deep generative models due to higher precision (higher COV-P, Fig. \ref{fig:precision_recall}), where the diffusion model generates higher quality synthesis routes. Interestingly, $\texttt{DiffSyn}$ achieves the lowest mean absolute error (MAE) for 10/12 synthesis parameters (Fig. \ref{fig:fig2}c), despite not being explicitly trained on the MAE objective like the regression-based models.

\subsubsection{Rationalizing the superior performance of generative approaches}
\label{sec:rationalizing_generative_models}
For a given target material, there are a range of possible values for each synthesis parameter (e.g., temperature) at which a material can be synthesized (i.e. \textit{synthesis window}). We show the distribution of crystallization times from published syntheses (grey, Fig. \ref{fig:fig2}d).

This distribution of synthesis parameters renders structure-synthesis relationship \textit{one-to-many} instead of one-to-one. Regression-based AMD is deterministic, and thus outputs a point prediction that comes from a weighted average of the distribution (Fig. \ref{fig:fig2}d). Generative models like GAN also output a point prediction as they often suffer from mode collapse \cite{zhang2018convergence}. Although generative models such as NF and VAE address mode collapse, they lack expressivity and fail to accurately capture the ground-truth literature distribution. In contrast, \texttt{DiffSyn} captures the ground-truth distribution. Furthermore, we compare the predicted and true joint distributions of \textit{multiple} synthesis parameters for all of these approaches for the AEL zeolite (Fig. \ref{fig:2d_dist}). Only deep generative approaches (NF, VAE, and \texttt{DiffSyn}) capture the ground-truth joint distribution of crystallization temperatures and times for the AEL structure, with \texttt{DiffSyn} most accurately capturing that joint distribution. \texttt{DiffSyn} captures most of the ground truth points, including some outliers; however, \texttt{DiffSyn} fails to predict points in a minor mode (bottom right of Guided Diffusion panel in Fig. \ref{fig:2d_dist}), which are extreme outliers (e.g., low crystallization temperature and long crystallization time).

The distribution of synthesis parameters is also \textit{multi-modal}. We plot the PCA of all synthesis parameters for the aforementioned AEL structure in Fig. \ref{fig:2d_pca_compare_with_struct}, which shows that the true distribution has multiple modes---2 in this case. Regression-based models (AMD, BNN) predict only one of the modes. GMM predicts synthesis routes that are far out of distribution. GAN suffers from mode collapse to one of the modes. NF and VAE capture both modes, but also generate a large number of false positives. This behavior arises from the low expressivity of VAEs and NFs, which use one-step decoding and affine invertible layers, respectively. In contrast, \texttt{DiffSyn} accurately predicts the true distribution because it generates high-quality and diverse outputs. Consequently, \texttt{DiffSyn} generates synthesis routes that overlap with unseen literature-reported synthesis parameters (Fig. \ref{fig:fig2}e). For a discussion on diversity of generated samples, please refer to Supplementary Section \ref{sec:generalization_diversity}.

\subsubsection{Learning chemically meaningful relationships}
We perform an unsupervised, hierarchical clustering of zeolite structures based on their learned representations (Fig. \ref{fig:fig3}a), and observe distinct clusters according to their corresponding structural features (e.g., number of channels, largest free sphere diameter). The clustering indicates that the zeolite encoder has learned to separate structurally distinct materials. Consequently, the chemical guidance (Fig. \ref{fig:fig1}b), which requires learning good representations of the zeolite and OSDA, guides the generative process towards the desired target material. 

Given that \texttt{DiffSyn} learns the joint distribution of multiple synthesis parameters, we inspect two synthesis parameters (crystallization temperature and time) in Fig. \ref{fig:physical_relationship} for two unseen zeolite-OSDA systems. An inverse relationship is observed between generated temperatures and times. This observation aligns well with the Arrhenius equation
\begin{equation}
    \label{eqn:arrhenius}
    k = Ae^{\frac{-E_a}{RT}},
\end{equation}
where crystallization time (related to rate $k$) is inversely related to temperature $T$. In addition, generated H$_2$O/T (here, T refers to tetrahedral framework atoms or T-atoms, rather than temperature) and framework density (FD$_{\text{Si}}$) of zeolite structure correlate positively (Spearman's coefficient: 0.673; Fig. \ref{fig:h2o_vs_fwd}). This finding agrees with Villaescusa's rule \cite{camblor1999synthesis}, which states that denser phases (higher FD$_{\text{Si}}$) are favored at lower concentrations of T-atoms (higher H$_2$O/T), showing that \texttt{DiffSyn} has learned domain-specific rules in materials synthesis. The model predictions also follow the thermodynamics of zeolite formation, where the generated crystallization temperature and framework density (FD$_{\text{Si}}$) of zeolite structure positively correlate (Spearman's coefficient: 0.931; Fig. \ref{fig:temp_vs_fwd}). This finding agrees with the thermodynamic argument from Ostwald’s rule of stages, which states that higher temperatures enable the synthesis to overcome the activation barrier to form more stable structures with higher framework densities \cite{le2019process, pan2024zeosyn}.

\subsection{Case studies} \label{case_studies} 
\begin{figure}[H]
    \centering
    \includegraphics[width=\linewidth]{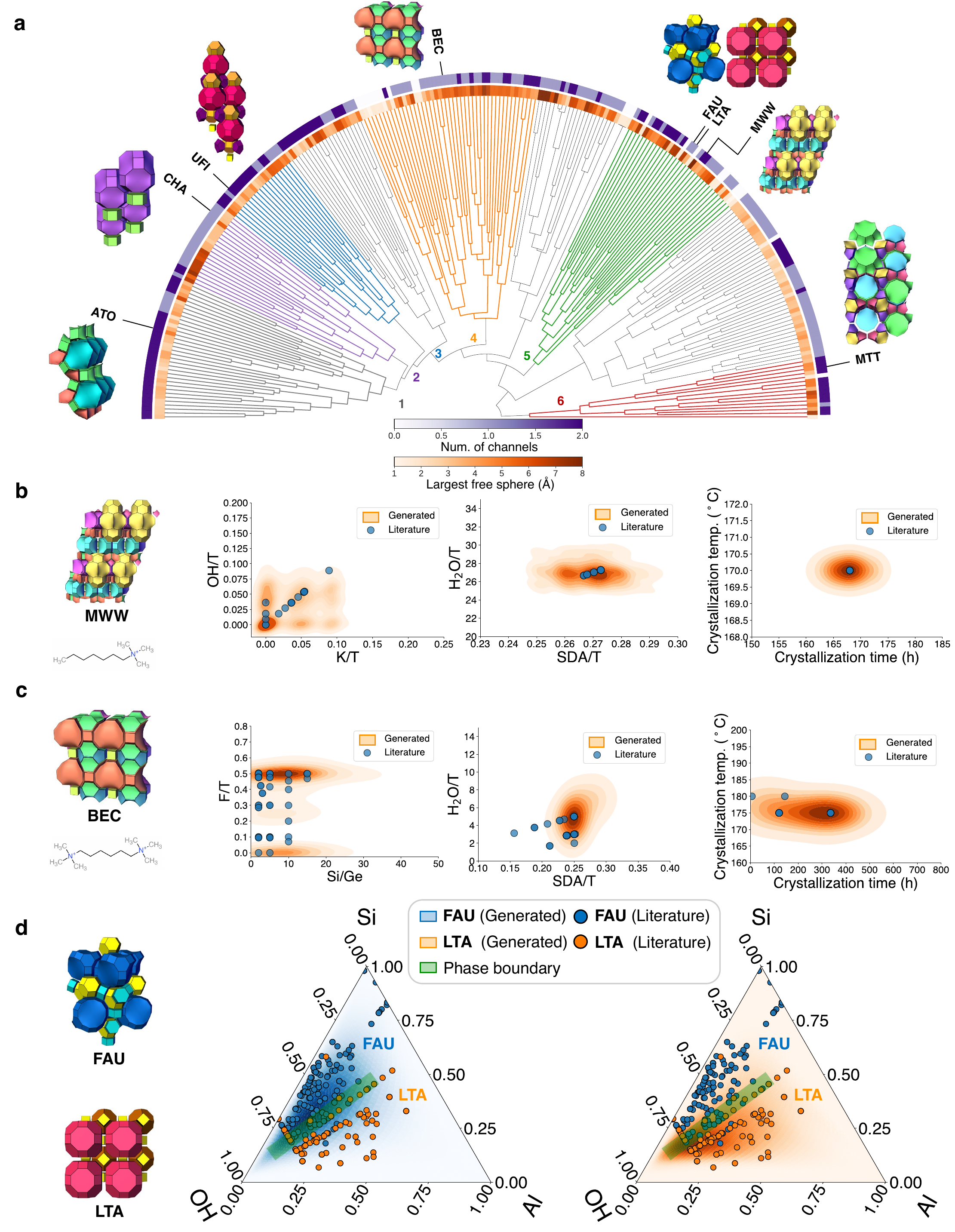}
    \caption{\textbf{Case studies on unseen materials systems}. 
    \textbf{(a)} Hierarchical clustering of zeolite structures. Splits are obtained via agglomerative hierarchical clustering of learned zeolite representations. Each structure (with a 3-letter code as its name) is colored by its number of channels (NC) in purple, and its largest free sphere (LFS) in orange. This leads to several distinct clusters: 
    (1) high NC, low LFS;
    (2) low NC, low+high LFS;
    (3) high NC, low+high LFS;
    (4) low NC, high LFS;
    (5) low+high NC, high LFS;
    (6) high NC, high LFS.
    Generated synthesis routes for unseen materials systems \textbf{(b)} MWW structure (in cluster 5) with N,N,N-trimethylhexan-1-aminium as OSDA. \textbf{(c)} BEC structure (in cluster 4) with pentane-1,5-diyl-bis(trimethylammonium) as OSDA. Orange heatmaps refer to synthesis routes generated by \texttt{DiffSyn}, while blue points refer to literature-reported synthesis routes.
    \textbf{(d)} Competing phases FAU and LTA (in cluster 5). Heatmaps refer to generated routes, while points refer to literature-reported synthesis routes. Notice that the model accurately predicts the phase boundary (green shaded region) between FAU and LTA.
    }
    \label{fig:fig3}
\end{figure}

We compare \texttt{DiffSyn}-generated synthesis routes to literature-reported synthesis routes for diverse zeolite-OSDA systems that are synthetically interesting and industrially useful. The generated routes for these unseen systems (MWW, BEC, and a pair of competing phases---FAU and LTA) provide evidence of \texttt{DiffSyn} learning meaningful synthesis-structure relationships (Fig. \ref{fig:fig3}). Other systems (MTT and ATO) are analyzed in the SI (Fig. \ref{fig:CS_MTT} and \ref{fig:CS_ATO}, respectively). 
 
We first consider the MWW phase, a unique two-dimensional structure with 10-membered rings and large cavities, with applications including isomerization \cite{corma1994proposed} and aromatization \cite{ma2001mo}. The generated OH$^-$/T, K$^+$/T, H$_2$O/T, SDA/T, and crystallization temperature/time overlap significantly with ground-truth synthesis parameters (Fig. \ref{fig:fig3}b). We also test the model on a significantly different structure: BEC, a large-pore zeolite. BEC has a three-dimensional pore topology with intersecting 12-membered ring channels, with applications in isomerization \cite{zhang2017single} and epoxidation \cite{moliner2008synthesis}. \texttt{DiffSyn}-generated synthesis parameters closely agree with synthesis parameters reported in the literature, particularly Si/Ge, F$^-$/T, and crystallization temperature/ time (Fig. \ref{fig:fig3}c). This prediction aligns with reports that Ge and F$^-$ stabilize the \textit{d4r} composite building unit of the BEC structure during synthesis \cite{villaescusa2007pure}. This finding suggests that \texttt{DiffSyn} learns how particular heteroatoms or synthesis conditions favor the formation of specific building units within zeolites. However, the generated synthesis parameters do not always fully recall the ground-truth for BEC synthesis. For example, the model fails to predict the full range of possible SDA/T values that have been identified in past recipes (SDA/T = 0.15--0.25; Fig. \ref{fig:fig3}c). 

Typically, materials synthesis aims to produce a single framework; however, if two phases during synthesis compete, some recipes result in two or more phases. Here, we applied \texttt{DiffSyn} to predict OSDA-free synthesis routes for FAU and LTA zeolites \cite{oleksiak2017organic}. Synthesis routes generated by \texttt{DiffSyn} align closely with literature-reported recipes (Fig. \ref{fig:fig3}d). Notably, \texttt{DiffSyn} accurately predicts the phase boundary region (green) between FAU and LTA in OSDA-free conditions, delineating the synthesis space under which the competing phases form. This result shows that \texttt{DiffSyn} not only accurately captures the forward relationship (structure-synthesis), but also the decision boundaries of the \textit{inverse} relationship (synthesis-structure), hence demonstrating its potential to enable phase-selective synthesis. Similarly, we demonstrate this capability for another pair of competing phases (ERI and KFI) in Fig. \ref{fig:ERI_KFI}. Taken together, these case studies illustrate \texttt{DiffSyn}'s ability to generalize to a variety of zeolite frameworks and their corresponding chemistries. 

\begin{figure}[t]
    \centering
    \includegraphics[width=\linewidth]{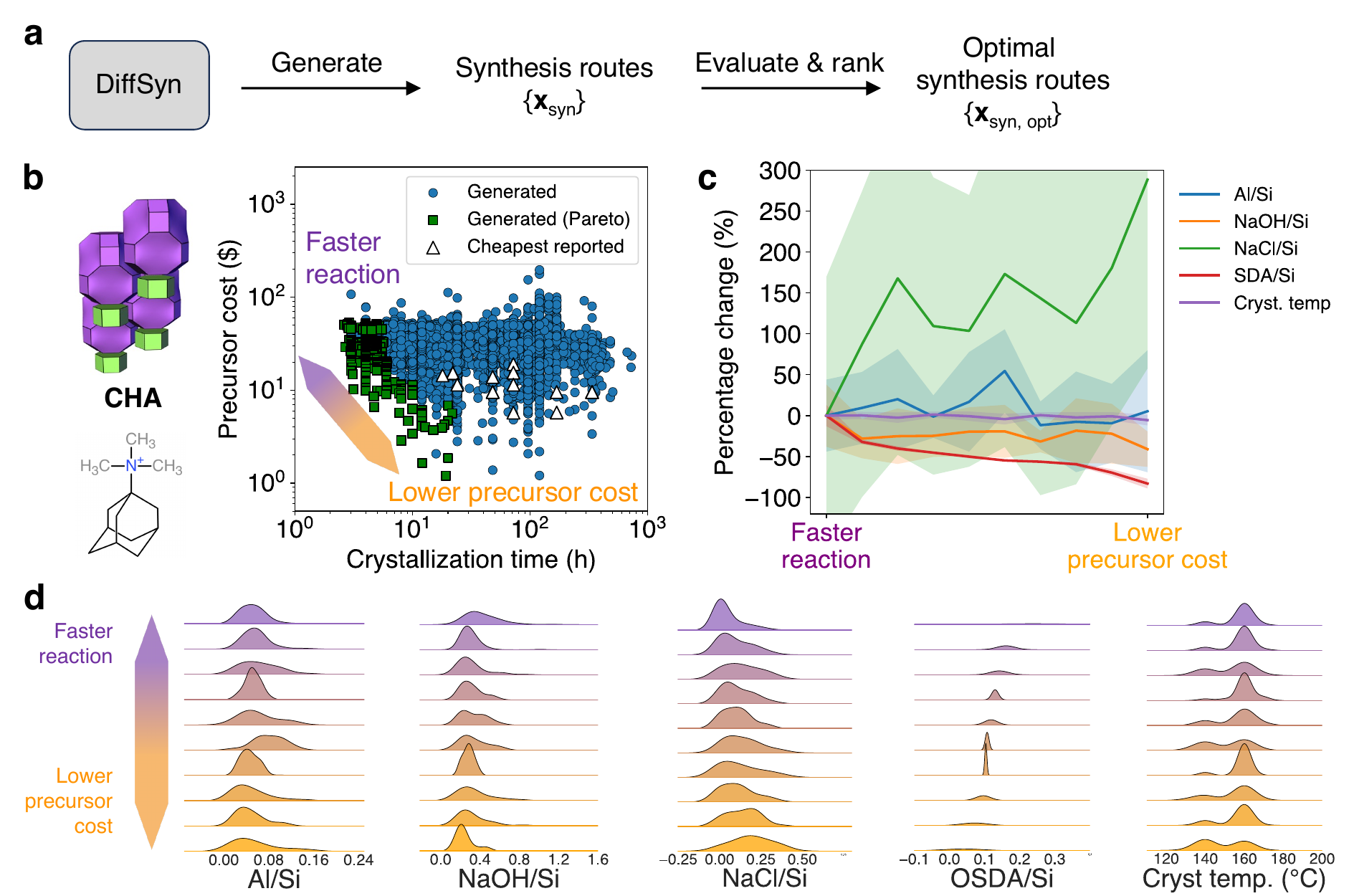}
    \caption{\textbf{Generating optimal synthesis routes}. \textbf{(a)} \texttt{DiffSyn} generates synthesis routes $\{\textbf{x}_{\text{syn}}\}$, which are then evaluated based on some metric of interest (cost, crystallization time) and ranked to find an optimal synthesis route. \textbf{(b)} Precursor cost and crystallization time of synthesis routes for the CHA structure (in cluster 2 of Fig. \ref{fig:fig3}a) with TMAda as the OSDA. Circles refer to \texttt{DiffSyn}-generated routes, triangles refer to 20 cheapest routes reported in literature. Notice the Pareto front (green squares) trades off cost and time. \textbf{(c)} Percentage change in generated synthesis parameters along the Pareto front. Solid lines and shaded regions refer to mean and standard deviation of the synthesis parameter, respectively. \textbf{(d)} Distributions of generated synthesis parameters along the Pareto front.
    }
    \label{fig:fig4}
\end{figure}

\subsection{Generating optimal synthesis routes} \label{sec:optimal}
We employ \texttt{DiffSyn} to generate more feasible synthesis routes (Fig. \ref{fig:fig4}a), by evaluating and ranking the generated synthesis routes based on precursor cost and crystallization time (see Table \ref{precursor_cost} for details). We apply this approach to the synthesis of CHA, using trimethyladamantyl ammonium (TMAda) as the OSDA \cite{di2016controlling} and generate an ensemble of synthesis routes with corresponding precursor costs and crystallization times (Fig. \ref{fig:fig4}b). Among the generated routes, there are Pareto optimal routes that trade off low precursor cost and fast synthesis time. These two objectives are optimal in different regions in the synthesis space---the OSDAs that produce CHA crystals faster tend to be more expensive (Fig. \ref{fig:pca_cost_time}). Some of these Pareto-optimal routes generated by \texttt{DiffSyn} have lower crystallization time and lower precursor cost compared to the 20 least expensive synthesis routes reported in literature. 

We inspect the evolution of synthesis parameters on the Pareto front. Notably, Al/Si and NaOH/Si remain relatively constant (Fig. \ref{fig:fig4}c). In contrast, SDA/Si decreases as we prioritize lower cost over faster reaction as the OSDA typically drives precursor cost. The distributions of generated synthesis parameters also change along the Pareto front (Fig. \ref{fig:fig4}d). This analysis reveals that raising the crystallization temperature from 140 to 160 $^\circ \text{C}$ while simultaneously increasing NaOH/Si and OSDA/Si would favor faster reaction (Fig. \ref{fig:fig4}d). This assessment shows how varying the joint distributions of synthesis parameters can accelerate synthesis (synthesizing desired structure in shorter time at higher temperature).

\subsection{Experimental and DFT validation}
\begin{figure}[t]
    \centering
    \includegraphics[width=\linewidth]{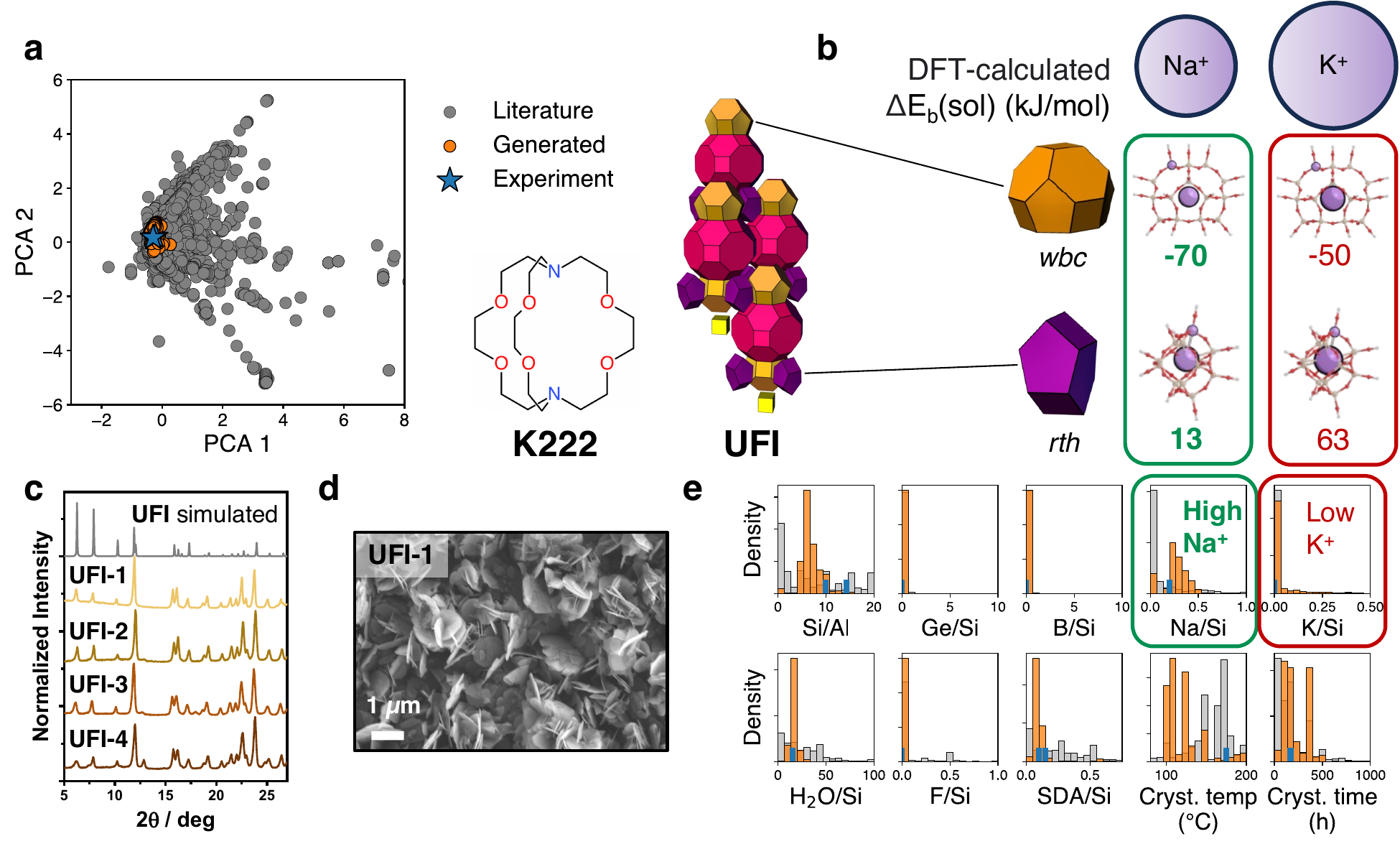}
    \caption{\textbf{Experimental and DFT validation.} We leverage \texttt{DiffSyn} to generate synthesis routes for the UFI structure, which resulted in successful crystallization of the desired structure \textbf{(a)} PCA of synthesis parameters of all routes reported literature (grey), \texttt{DiffSyn}-generated for UFI (orange) and experimentally-verified (blue) in this work. \textbf{(b)} DFT-calculated binding energies ($\Delta E_b$) between cations (Na$^+$, K$^+$) and building units of UFI (\textit{rth}, \textit{wbc}). Notice Na$^+$ has stronger (more negative) $\Delta E_b$ than K$^+$, which suggests that Na$^+$ (compared to K$^+$) favors UFI formation. This rationalizes why \texttt{DiffSyn} suggested synthesis routes with high Na$^+$/Si and low K$^+$/Si for UFI (in \textbf{(e)}) . \textbf{(c)} Simulated and measured XRD patterns of the synthesized UFI samples. \textbf{(d)} SEM image of UFI. \textbf{(e)} Distributions of individual synthesis parameters. Blue are experimentally-verified synthesis routes, orange is \texttt{DiffSyn}-generated, and grey is all literature-reported routes.
    }
    \label{fig:fig5}
\end{figure}

We validate \texttt{DiffSyn} by experimentally synthesizing a UFI zeolite from recipes generated by our model with Kryptofix 222 (K222) as the OSDA (Fig. \ref{fig:fig5}) \cite{lee2025data}. UFI has potential applications in industrially relevant reactions (e.g., selective catalytic reduction of NO$_x$ \cite{jo2016synthesis}). The UFI-K222 system has not been reported in prior literature and is not present in the training dataset, and hence serves as a test of \texttt{DiffSyn}'s out-of-distribution generalization. 

Fig. \ref{fig:fig5}a shows the PCA of 1000 \texttt{DiffSyn}-generated synthesis routes for UFI (orange), which constitutes a subspace of literature-reported synthesis routes for all zeolites (grey). We retrieve the $k$-nearest neighbors ($k=5$) of generated UFI synthesis from the literature-reported synthesis routes (Fig. \ref{fig:pca_knn}). Among the retrieved synthesis, the top 2 most similar frameworks (PAU and RHO) share a common \textit{lta} composite building unit (CBU) with UFI (Fig. \ref{fig:knn_zeo_freq}). Beyond this, the remaining frameworks do not share any common CBUs with UFI. This observation is supported by a previous work \cite{schwalbe2019graph} reporting that majority of competing phases do not share any CBUs, suggesting that structure-synthesis relationships are complex and cannot be rationalized with building units alone. 

We visualize individual synthesis parameters of generated routes in Fig. \ref{fig:fig5}e (orange histograms), including gel compositions (Si/Al, Ge/Si, B/Si, Na$^+$/Si, K$^+$/Si, H$_2$O/Si, F$^-$/Si) and reaction conditions (crystallization temperature and time). For inorganic cations, \texttt{DiffSyn} recommends high Na$^+$/Si and low K$^+$/Si for UFI synthesis (Fig. \ref{fig:fig5}e). We rationalize this observation by calculating binding energies ($\Delta E_b$) of inorganic cations (Na$^+$ and K$^+$) in building units of UFI (\textit{wbc} and \textit{rth}) using density functional theory (DFT) as shown in Fig. \ref{fig:fig5}b (details in Section \ref{sec:dft}). The calculations reveal that Na$^+$ binds more strongly to \textit{wbc} and \textit{rth} ($\Delta E_b^{sol}$ = --70 and +13 kJ/mol) compared to K$^+$ ($\Delta E_b^{sol}$ = --50 and +63 kJ/mol), suggesting that Na$^+$ ions stabilize building units of UFI better than K$^+$. These binding energies explain the high Na$^+$/Si and low K$^+$/Si in \texttt{DiffSyn}-generated routes.

We experimentally test the \texttt{DiffSyn}-generated routes for UFI, resulting in four successful experimental syntheses of the UFI material (Fig. \ref{fig:fig5}a, blue). Powder X-ray diffraction (XRD) patterns for synthesized UFI samples closely match the simulated diffraction pattern (Fig. \ref{fig:fig5}c), confirming that the crystallized material has the UFI structure. The resultant crystals had a measured Si/Al$_{\text{ICP}}$ of 19.0, the highest yet recorded in UFI synthesis (see Section \ref{icp} for details on Si/Al$_{\text{ICP}}$). This composition is important because higher Si/Al$_{\text{ICP}}$ is associated with higher thermal stability under catalytic conditions \cite{cruciani2006zeolites}. Additionally, the crystals exhibit a house-of-cards morphology (Fig. \ref{fig:fig5}d), making them highly promising catalysts for a diverse range of applications. 

While the experimentally verified synthesis routes fall within the generated distributions, the model generated unusually low crystallization temperatures with a major mode at 100--150$^\circ \text{C}$ (Fig. \ref{fig:fig5}e). The generated minor mode at 175$^\circ \text{C}$ was chosen based on domain expertise because higher temperatures accelerate crystallization kinetics and higher temperatures ($>$170$^\circ \text{C}$) avoid the formation of the undesired LTA competing phase. The higher temperature allowed us to use a relatively shorter crystallization time of 7 days (168 h). These considerations underscore a powerful synergy between model and human expertise to achieve the desired synthesis outcome. Nonetheless, \texttt{DiffSyn} generated 9/10 synthesis parameters as the major mode. The successful synthesis of UFI, enabled by \texttt{DiffSyn}, is a testament to \texttt{DiffSyn} learning the intricacies of zeolite synthesis without explicit chemical encodings and highlights its ability to recommend suitable synthesis parameters for unseen materials systems.

\section{Discussion}
Zeolite synthesis is a complex task with a high-dimensional space where synthesis parameters interact. This complexity underscores the need for generative models that capture the intricacies of materials synthesis and circumvent expensive first-principles approaches. By capturing the one-to-many and multi-modal nature of structure-synthesis relationships, \texttt{DiffSyn} outperforms previous methods for zeolite synthesis prediction.

\texttt{DiffSyn} requires the OSDA to be known \textit{a priori}. Including the OSDA as an input ensures the model captures the influence of the organic template. Recent work on OSDA design has enabled the selection of suitable OSDA inputs into \texttt{DiffSyn} \cite{xie2025comprehensive, schwalbe2021priori, hoffman2024learning}. The combination of OSDA design tools and \texttt{DiffSyn} could enable an end-to-end predictive workflow for zeolite synthesis. While our work focuses solely on continuous variables, the synthesis outcome can also be influenced by categorical/discrete variables, such as the precursor choice \cite{he2023precursor, prein2025retro, noh2025retrieval} and presence of seed crystals. These discrete variables present a compelling opportunity for future work in ML for synthesis modeling---e.g., discrete diffusion \cite{lou2023discrete}, autoregressive models \cite{kim2024large, okabe2024large, prein2025language, song2025accurate}, reinforcement learning \cite{karpovich2024deep, pan2022deep} for sampling discrete synthesis parameters.

For diffusion models, the inference speed is slower than other generative approaches due to sequential denoising. However, since the bottleneck of materials synthesis is experimental synthesis (e.g., reaction time of weeks), inference speed is not a significant problem. Even if sampling speed becomes a bottleneck, techniques such as denoising diffusion implicit model (DDIM) sampling \cite{song2020denoising} can trade off sampling quality for faster sampling. Flow matching models \cite{lipman2022flow} could be an alternative to balance quality and speed. Notably, training \texttt{DiffSyn} requires optimizing hyperparameters for classifier-free guidance such as $p_{\text{uncond}}$ and $w$ as shown in Fig. \ref{fig:p_uncond_and_cond_scale}, which may be challenging if compute is limited.

The successful synthesis of UFI enabled by \texttt{DiffSyn} exemplifies the potential of our approach to guide experimental zeolite synthesis. However, a comprehensive validation of \texttt{DiffSyn} would ideally also demonstrate that recipes substantially diverging from the generated ones fail to yield the target phase; such an evaluation is beyond the scope of this study and is reserved for future work. We hypothesize diffusion models could be applied to other materials systems, particularly when data are plentiful. This approach signifies a shift from regression to generative models for materials synthesis, where the latter is needed to effectively model one-to-many structure-synthesis relationships. This work has enabled the prediction of probable and diverse synthesis pathways of microporous materials, signifying a step toward bridging computational materials design (\textit{what} to synthesize) and synthesis planning (\textit{how} to synthesize).

\section{Methods}
\subsection{Zeolite and OSDA representations} 
\label{featurization}
\textbf{Zeolite} We adopt two different approaches of encoding the zeolite structure. First, invariant features (e.g., ring sizes, largest included sphere) are retrieved from the International Zeolite Association (IZA) database \cite{baerlocher2008database}. These features serve as inputs into a MLP encoder. Second, an equivariant graph neural network (EGNN) \cite{satorras2021n} encodes the zeolite as a graph (see Section \ref{model_implementation}). \textbf{OSDA} We optimized each OSDA in the gas phase with the MMFF94 force ﬁeld from RDKit-generated conformers \cite{halgren1999mmff}. Each OSDA is featurized using its physicochemical descriptors (e.g., molecular volume and 2D shape descriptors) averaged over all conformers \cite{schwalbe2021priori, jensen2021discovering}. The OSDA features are defined in Table \ref{osda_descriptors}. The zeolite embedding is concatenated with the OSDA embedding and further encoded using the fusion encoder before the joint embedding guides the reverse diffusion process to generate synthesis routes (Fig. \ref{fig:fig1}). Model performance across the denoising diffusion trajectory can be found in Fig. \ref{fig:metrics_vs_t}.

\subsection{Denoising diffusion probabilistic models} \label{ddpm}
Denoising diffusion probabilistic models (DDPMs) \cite{ho2020denoising} are generative models that use a diffusion process to generate data by reversing a forward process that incrementally adds noise to the data. The forward process gradually corrupts the data $x_0$ into a noisy sample $x_t$ by adding Gaussian noise in a Markov chain:
\begin{equation}
q\left(x_t \mid x_{t-1}\right)=\mathcal{N}\left(x_t ; \sqrt{1-\beta_t} x_{t-1}, \beta_t \mathbf{I}\right)
\end{equation}
where $\beta_t$ is the variance at timestep $t$. It is often useful to sample $x_t$ directly from $x_0$:
\begin{equation}
q\left(x_t \mid x_0\right)=\mathcal{N}\left(x_t ; \sqrt{\bar{\alpha}_t} x_0,\left(1-\bar{\alpha}_t\right) \mathbf{I}\right)
\end{equation}
where $\alpha_t=1-\beta_t, \quad \bar{\alpha}_t=\prod_{s=1}^t \alpha_s$. The generative process learns to reverse this corruption step by step (Fig. \ref{fig:fig1}b), sampling from a distribution $p_\theta\left(x_{t-1} \mid x_t\right)$ parameterized by a neural network. This reverse process can also be represented as a Gaussian distribution
\begin{equation}
p_\theta\left(x_{t-1} \mid x_t\right)=\mathcal{N}\left(x_{t-1} ; \mu_\theta\left(x_t, t\right), \Sigma_\theta\left(x_t, t\right)\right)
\end{equation}
Here, we learn the mean while fixing the variance as
\begin{equation}
\mu_\theta\left(x_t, t\right)=\frac{1}{\sqrt{\alpha_t}}\left(x_t-\frac{\beta_t}{\sqrt{1-\bar{\alpha}_t}} \epsilon_\theta\left(x_t, t\right)\right), \quad \Sigma_\theta\left(x_t, t\right)=\beta_t \mathbf{I}
\end{equation}
where $\epsilon_\theta\left(x_t, t\right)$ is a neural network trained to predict the noise $\epsilon$. In this work, we use a U-Net (Fig. \ref{fig:fig1}c) \cite{ronneberger2015u}. The training objective is to minimize a variational bound $\mathbb{E}_{t, x_0, \epsilon}\left[\left\|\epsilon-\epsilon_\theta\left(x_t, t\right)\right\|^2\right]$, which results in the diffusion model
\begin{equation}
p_\theta\left(x_0\right)=\int p\left(x_T\right) \prod_{t=1}^T p_\theta\left(x_{t-1} \mid x_t\right) d x_{1: T}
\end{equation}

\subsection{Chemically guided diffusion model} \label{sec:cfg}
In standard guided diffusion models, a classifier is used to guide the generation process by adjusting the score to steer the model towards specific target classes. In contrast, classifier-free guidance \cite{ho2022classifier} eliminates the need for a separate classifier by conditioning the diffusion model directly on the desired attributes. During training, the score function $\tilde{s}_{\theta}$ is trained both with and without conditioning $c$ using a null token $\varnothing$. This training is done by randomly setting $c$ to the unconditional null token $\varnothing$ with some probability $p_{\text{uncond}}$. Sampling is then performed using a linear combination of the conditional  and unconditional score estimates:

\begin{equation}
    \tilde{s}_\theta\left(x_t, t, c\right)=(1+w) s_\theta\left(x_t, t, c\right)-w s_\theta\left(x_t, t, \varnothing \right)
\label{eq:cfg}
\end{equation}
where $c$ refers to the chemical guidance from zeolite and OSDA embeddings shown in Fig. \ref{fig:fig1}, while $w$ is the strength of the chemical guidance. Supplementary Section \ref{model_implementation} contains the implementation details.

\subsection{Model evaluation}

\subsubsection{Metrics}
\label{section:metrics}
For each test zeolite-OSDA system, we sample 1000 synthesis routes using the model and compute the following metrics with reference to \textit{unseen} synthesis routes reported in literature. \textbf{Wasserstein distance} measures the distance between two probability distributions by finding the minimum cost to move probability mass from one distribution to another \cite{rubner1998metric}. The Wasserstein metric captures differences in both the location and the shape of distributions (including spread and the presence of multiple modes). While Kullback–Leibler (KL) divergence may be used, the Wasserstein distance is chosen as it fulfills all requirements of a metric: non-negativity, identity of indiscernibles, symmetry and triangle inequality. Moreover, a significant drawback of KL divergence is its behavior with distributions that do not have overlapping support. If there is any point where one distribution has a zero probability and the other has a non-zero probability, the KL divergence can become infinite or undefined. Wasserstein distance, in contrast, provides a meaningful and finite distance even for distributions with non-overlapping supports. Wasserstein distance has been widely adopted in generative modeling (e.g., WGANs \cite{arjovsky2017wasserstein} employ a form of Wasserstein distance) because it provides a smooth, sensitive measure of distributional differences. Small improvements in how well the model captures tails or secondary modes are reflected by a lower Wasserstein distance, guiding us during model development to favor settings that capture the full distribution. \textbf{Coverage} Inspired by Xu \cite{xie2021crystal}, we use two coverage metrics, COV-R (recall) and COV-P (precision), to measure the similarity between sets of generated and literature-reported synthesis for each zeolite-OSDA system. Intuitively, COV-R measures the
fraction of literature synthesis routes being correctly predicted, and COV-P measures the fraction of generated synthesis routes being probable. COV-F1 is computed as the harmonic mean of COV-R and COV-P. Refer to Supplementary Section \ref{coverage_metrics} for a detailed justification of the metrics.

\subsubsection{Baselines}
\label{section:baseline}
\textbf{Random} A random dummy baseline corresponds to picking a random point in synthesis space. \textbf{AMD} Schwalbe-Koda et al. proposed a deterministic regression-based approach using average minimum distance (AMD) for zeolite structural featurization for synthesis prediction task \cite{schwalbe2023inorganic}. \textbf{BNN} Bayesian neural networks \cite{jospin2022hands} extend standard neural networks by incorporating Bayesian inference by treating network weights as probability distributions, enabling a distribution of outputs. We implement a classical generative approach, Gaussian mixture model (\textbf{GMM}) \cite{reynolds2009gaussian}, which models data probabilistically as a sum of Gaussians (each with its mean and covariance). We also implement deep generative baselines: conditional variational autoencoder (\textbf{VAE}) \cite{karpovich2023interpretable, karpovich2021inorganic}, conditional generative adversarial network (\textbf{GAN}) \cite{mirza2014conditional} and conditional normalizing flow (\textbf{NF}). For NF, we use real-valued non-volume preserving (RealNVP) transformations \cite{dinh2016density}.

\subsection{Experimental methods}

\subsubsection{Synthesis of UFI}
We used colloidal silica (Ludox AS-40, 40 wt$\%$, Aldrich), aluminum hydroxide (Al(OH)$_3$, SPI Pharmacy), sodium hydroxide (NaOH, 50$\%$, Aldrich), with Kryptofix 222 (98$\%$, Ambeed) and tetramethylammonium hydroxide pentahydrate (TMAOH$\cdot$5H$_2$O, 97$\%$, Aldrich) as OSDAs to synthesize UFI. The OSDAs are selected by the method described in \cite{lee2025data}. The synthesis mixture had the molar composition 1.0 SiO$_2$, 0.033 Al$_2$O$_3$, 0.1 Na$_2$O, 15 H$_2$O, 0.15 Kryptofix 222, 0.025 TMA$_2$O. 0.021 g of aluminum hydroxide (Al(OH)$_3$) was dissolved in a solution containing 0.64 g of NaOH, 6.9 g of deionized water, 0.36 g of TMAOH$\cdot$5H$_2$O. After the solution became clear, 6 g of Ludox AS-40 and 2.31 g of Kryptofix 222 were added. 4 wt$\%$ of H-UFI seed material was included. The mixture was stirred and aged at room temperature for 24 h, then transferred to Teflon-lined 23 mL autoclaves and heated at 175 $^\circ\text{C}$ under dynamic conditions for 7 days. The resulting zeolites were recovered by centrifugation (12,000 rpm, 10 min), washed with deionized water three times, and dried overnight at 110 $^\circ\text{C}$. Samples were then heated at a ramp rate of 1 $^\circ\text{C}$/min to 580 $^\circ\text{C}$ in flowing dry air (100 mL/g of zeolite) for 6 h to remove OSDA molecules.

\subsubsection{X-ray diffraction}
Zeolite crystal structures were analyzed using powder X-ray diffraction (XRD) with a Bruker D8 diffractometer (Cu-K$\alpha$ radiation, $\lambda$ = 1.5418 $\text{\AA}$, 40 kV, 40 mA). Diffraction patterns were recorded over a 2$\theta$ range of 5$^\circ$–45$^\circ$, with an angular step size of 0.02$^\circ$ and a scanning rate of 4$^\circ$/min to confirm phase purity.

\subsubsection{Scanning electron microscopy}
The morphology of the calcined zeolite crystals was observed using a Zeiss Merlin High-Resolution scanning electron microscope (SEM). Zeolite samples were prepared as fine powders and mounted on carbon tape. Images were collected at 2.0 kV, 100 pA, and a working distance of 6.7 mm using the HE-SE2 detector in High-Resolution Column mode.

\subsubsection{Elemental analysis} \label{icp}
The elemental composition of silicon, and aluminum (Si/Al$_{\text{ICP}}$) of synthesized UFI zeolites was determined using inductively coupled plasma atomic emission spectroscopy (ICP-AES, Agilent 5100). Si/Al$_{\text{ICP}}$ is an important property of zeolites as high Si/Al$_{\text{ICP}}$ is correlated with high thermal stability \cite{cruciani2006zeolites}. Samples (10 mg) were digested in 100 $\mu$L of hydrofluoric acid (48 wt$\%$, Sigma-Aldrich) for 24 hours, followed by dilution to 10 g with 2 wt$\%$ aqueous nitric acid (GFS Chemicals). Calibration curves were constructed with six-point standards using ICP solutions of 1,000 ppm Si, Al, and Na in 2 wt$\%$ HNO$_3$ (Sigma-Aldrich, TraceCERT).

\subsection{DFT methods} \label{sec:dft}

Density functional theory (DFT) calculations were conducted using the ORCA package (v5.0.4) \cite{neese_orca_2012, neese_software_2018}. For all calculations, the $\omega$B97X-D hybrid functional \cite{chai_long-range_2008, chai_systematic_2008} was employed along with the def2 triple-$\zeta$ basis set with polarization (def2-TZVP) \cite{weigend_balanced_2005}. 
During electronic optimization, wavefunctions were converged when energy changes were less than 10$^{-8}$ Ha. Geometry optimizations were performed until the energy varied by less than 5×10$^{-6}$ Ha and the forces on all atoms were below 3×10$^{-4}$ Ha bohr$^{-1}$, adhering to the default ORCA optimization protocol. Where indicated, the SMD solvation model was applied to estimate solvation effects on the energies of each structure, using water as the solvent with a dielectric constant ($\epsilon$) of 80.4 \cite{marenich_universal_2009}.

Binding energies were calculated relative to an isolated ion in vacuum and an empty Al-substituted composite building units (CBUs):
\begin{equation}
    \Delta E_b=E[\mathrm{M} \bullet \mathrm{CBU}]-E\left[\mathrm{M}^{n+}\right]-E\left[\mathrm{CBU}^{n-}\right]
\end{equation}
where $E[\mathrm{M} \bullet \mathrm{CBU}]$ represents the energy of the metal ion positioned within the Al-substituted CBU, $E\left[\mathrm{M}^{n+}\right]$ is the energy of the isolated ion, and $E\left[\mathrm{CBU}^{n-}\right]$ is the energy of the empty Al-substituted CBU containing $n$ Al atoms. When CBUs offered multiple potential single Al positions or several configurations of two Al atoms, we selected the $\Delta E_b$ corresponding to the most stable Al arrangement, as this stability likely reflects the thermodynamically preferred configuration. Given adequate synthesis time, Al atoms tend to occupy stable framework positions, as demonstrated in recent CHA studies \cite{lee_evolution_2022}. Atoms within CBUs, including terminal hydroxyl groups, were kept fixed during calculations. Energies for empty, Al-substituted CBUs and isolated ions (in vacuum or with implicit solvation) were determined via single-point calculations in ORCA with appropriate charges.

The optimization of the metal-docked ion within the CBU allowed movement of the ion but maintained fixed positions for the CBU atoms, replicating the constraints imposed by a crystal environment where surrounding framework atoms would restrict movement; unconstrained optimization could lead to significant atomic displacements in these molecular forms. This study aims to understand how these ions influence specific CBU structures, given that the OSDA likely directs \textit{lta} cage formation in these zeolites. Thus, CBU atoms were fixed to preserve a shape close to their zeolite-based structures, and binding energies were calculated with and without solvation effects from water.

Zeolite crystal structures for materials synthesized in this work were obtained from the International Zeolite Association (IZA) database \cite{baerlocher2008database}. Molecular models of individual CBUs for each synthesized framework were extracted by isolating them from their crystalline counterparts, with frameworks and corresponding CBUs. CBUs were derived from selected zeolite frameworks, as listed in the IZA database. Terminal SiOH groups were added to undercoordinated Si atoms to maintain tetrahedral coordination. These SiOH groups were oriented with Si–O–H bond angles of 180° to minimize hydrogen bonding (H-bonding) interactions between nearby SiOH groups 
(Fig. \ref{fig:dft_docking}).

\section{Data and code availability}
The corresponding code and dataset will be available at:
\url{https://github.com/eltonpan/zeosyn_gen}

\section{Acknowledgments}

The authors acknowledge funding from the Generalitat Valenciana through the Prometeo Program (CIPROM/2023/34), CSIC through the I-link+ Program (ILINK24035) and MIT International Science and Technology Initiatives (MISTI) Seed Funds between MIT and CSIC.
The authors also acknowledge partial funding from the National Science Foundation DMREF Awards 1922090, 1922311, 1922372, the Office of Naval Research (ONR) under contract N00014-20-1-2280, Kwanjeong Educational Fellowship, ExxonMobil, and the Agency for Science, Technology and Research. The authors thank Di (Daniel) Du, Allen Burton, Xiaochen Du, Soojung Yang and Serena Khoo for helpful discussions and suggestions.







\begin{appendices}

\pagebreak

\setcounter{page}{1}
\resetlinenumber


\begin{center}

{\LARGE Supplementary Information for}

{\LARGE \texttt{DiffSyn}: A Generative Diffusion Approach to Materials Synthesis Planning}

\vspace{0.5cm}

{\large Elton Pan$^1$, Soonhyoung Kwon$^2$, Sulin Liu$^1$, Mingrou Xie$^2$,
Alexander J. Hoffman$^1$, Yifei Duan$^1$, Thorben Prein$^3$,
Killian Sheriff$^1$, Yuriy Roman-Leshkov$^2$, Manuel Moliner$^4$,
Rafael Gomez-Bombarelli$^1$, Elsa Olivetti$^{1*}$}

\vspace{0.5cm}

$^1$ Department of Materials Science and Engineering, Massachusetts
Institute of Technology, Cambridge, MA, 02139, USA.

$^2$ Department of Chemical Engineering, Massachusetts Institute of
Technology, Cambridge, MA, 02139, USA.

$^3$ Department of Chemistry, Technische Universit\"at M\"unchen, M\"unchen,
80333, Germany.

$^4$ Instituto de Tecnolog\'{i}a Qu\'{i}mica, Universitat Polit\`{e}cnica de Val\`{e}ncia-Consejo Superior de Investigaciones Cient\'{i}ficas, Valencia,46022, Spain.

\end{center}

\affil[1]{\orgdiv{Department of Materials Science and Engineering}, \orgname{Massachusetts Institute of Technology}, \orgaddress{\city{Cambridge}, \state{MA}, \postcode{02139}, \country{USA}}}

\affil[2]{\orgdiv{Department of Chemical Engineering}, \orgname{Massachusetts Institute of Technology}, \orgaddress{\city{Cambridge}, \state{MA}, \postcode{02139}, \country{USA}}}

\affil[3]{\orgdiv{Department of Chemistry}, \orgname{Technische Universität München}, \orgaddress{\city{München}, \postcode{80333}, \country{Germany}}}

\affil[4]{\orgname{Instituto de Tecnolog\'{i}a Qu\'{i}mica, Universitat Polit\`{e}cnica de Val\`{e}ncia-Consejo Superior de Investigaciones Cient\'{i}ficas}, \orgaddress{\city{Valencia}, \postcode{46022}, \country{Spain}}}

\section{ZeoSyn dataset and preprocessing}
\label{zeosyn_dataset}
\textbf{Dataset} The ZeoSyn dataset \cite{pan2024zeosyn} contains comprehensive synthesis information on zeolites including gel composition, reaction conditions (crystallization time/temperature), precursors, and OSDAs. The dataset also includes the resulting zeolite structures formed (or lack thereof e.g., dense/amorphous phases) for each synthesis route. The dataset consists of 23,961 synthesis routes from 3,096 journal articles spanning the years 1966--2021. It contains data on 921 unique OSDA molecules, 233 zeolite topologies, and 1,022 unique materials. The gel compositions are a combination of 51 different gel components including Si, Al, P, Na$^+$, K$^+$, OH$^-$, F$^-$, Ge, Ti, B, Ga, V, OSDA, H$_2$O, and additional solvents. In the context of zeolite synthesis, the OSDA refers to an organic molecule that act as a 'template' that guides the arrangement of inorganic building blocks (e.g. Si, Al) to form the zeolite material.

\textbf{Dataset split} For data splits, a 80/20 train/test split via zeolite-OSDA systems. This is to prevent data leakage across systems and ensure the model generalizes to unseen chemical systems. This results in 1856 and 464 systems in the training and test sets, respectively. Furthermore, the training set is further partitioned via a 87.5/12.5 train/validation random split for hyperparameter tuning. Given most synthesis parameters do not follow the Gaussian distribution (see blue distributions in Fig. \ref{fig:fig2}e), we perform quantile transformation using \texttt{sklearn.preprocessing.QuantileTransformer} to map the raw synthesis parameters to a uniform distribution.

\section{Additional discussion on generalization and diversity of DiffSyn}
\label{sec:generalization_diversity}
\textbf{Generalization to completely out-of-distribution systems} Given the train/test split described in Appendix \ref{zeosyn_dataset}, it may be possible that \textit{either} the zeolite framework or the OSDA may be present in both train and test set. To more rigorously test \texttt{DiffSyn}, we evaluate \texttt{DiffSyn} on only test systems where \textit{both} zeolite and OSDA are not seen during training. On these strict out-of-distribution test systems that measures the true explorative power of DiffSyn, the Wasserstein distance metric (lower is better) is worse (0.45 for out-of-distribution test systems vs. 0.42 for original test systems). This is expected, due to the difficulty of these systems. However, this deterioration in performance is rather small. For reference, even if evaluated on this harder subset of test systems, \texttt{DiffSyn} still outperforms all previous methods (Fig. 2\ref{fig:fig2}a). This shows that \texttt{DiffSyn} is robust to out-of-distribution test systems.

\textbf{Diversity of generated samples} We observe that \texttt{DiffSyn} may place a higher density of generated samples on major modes, which results in the generated distribution having more spiky peaks compared to ground truth distribution (Fig. \ref{fig:fig2}e). This is because the data on which \texttt{DiffSyn} is trained on (ZeoSyn dataset) is mined from the scientific papers, which is prone to anthropogenic factors such as researchers reporting the most common/convenient value (e.g., crystallization temperature of 150$^\circ$C instead of 153$^\circ$C). We observe this “spiky peak” phenomenon for synthesis parameters such as crystallization temperature. However, while this phenomenon may decrease the diversity of generated samples, it does not entail mode collapse. For example, \texttt{DiffSyn} does not suffer the mode collapse issue unlike conditional GAN, while most accurately capturing the ground truth distribution compared to other generative approaches such as VAEs and normalizing flows (Fig. \ref{fig:2d_dist}). Moreover, additional case studies in Fig. \ref{fig:fig3}b, \ref{fig:fig3}b, \ref{fig:fig3}d, \ref{fig:CS_ATO}, \ref{fig:CS_MTT} all show that mode collapse is not a major problem, with \texttt{DiffSyn} maintaining output diversity. We do acknowledge that, however, reduced diversity may impede \texttt{DiffSyn}’s ability to capture minor modes, which have been discussed in the main text in Sections \ref{sec:rationalizing_generative_models} and \ref{case_studies} for AEL and BEC, respectively. 




\renewcommand{\thefigure}{A\arabic{figure}}
\renewcommand{\theHfigure}{A\arabic{figure}}
\setcounter{figure}{0}

\renewcommand{\thetable}{A\arabic{table}}
\renewcommand{\theHtable}{A\arabic{table}}
\setcounter{table}{0}

\begin{figure}[h]
    \centering
    \includegraphics[width=\linewidth]{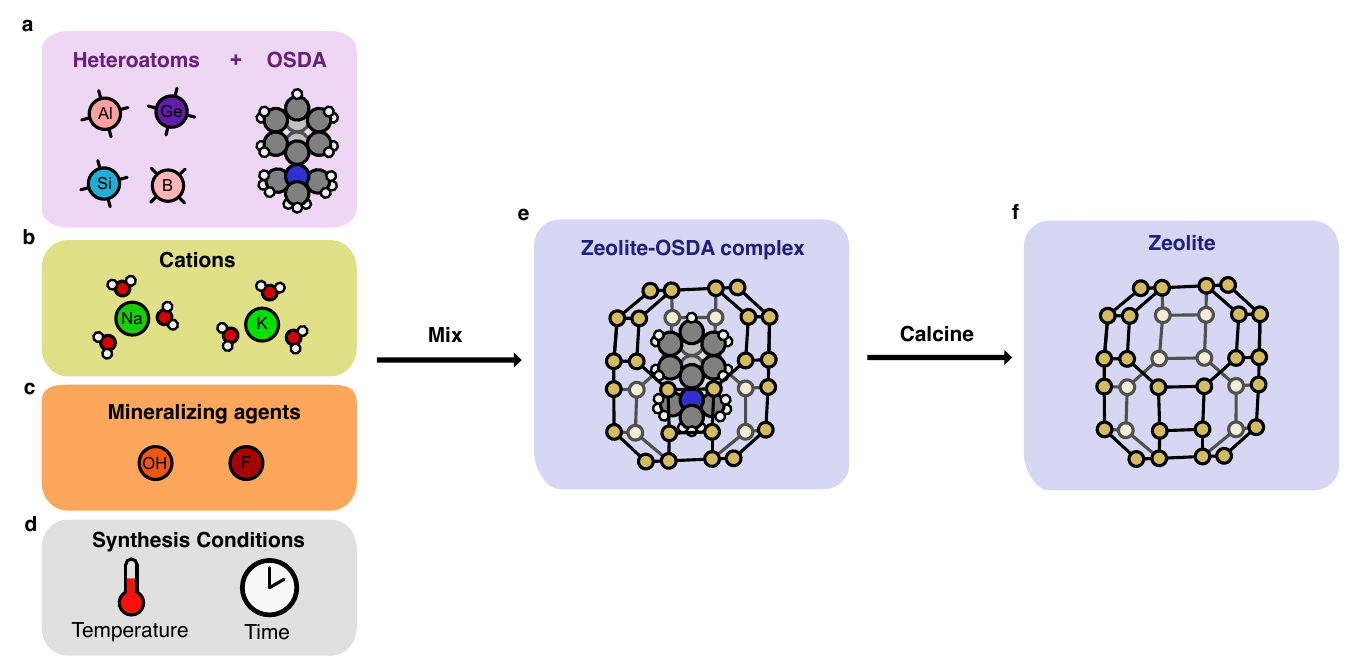}
    \caption{\textbf{Complexity of zeolite synthesis.} The synthesis of zeolite involves (a) heteroatoms and organic structure-directing agent (OSDA) (b) inorganic cations (c) mineralizing agents, mixed under (d) synthesis conditions to form (e) zeolite-OSDA complex, where the OSDA "templates" the resulting zeolite's pore structure. Calcination then removes the OSDA to yield the (f) zeolite structure.
    }
    \label{fig:osda_templating}
\end{figure}

\begin{figure}[h!]
    \centering
    \includegraphics[width=\linewidth]{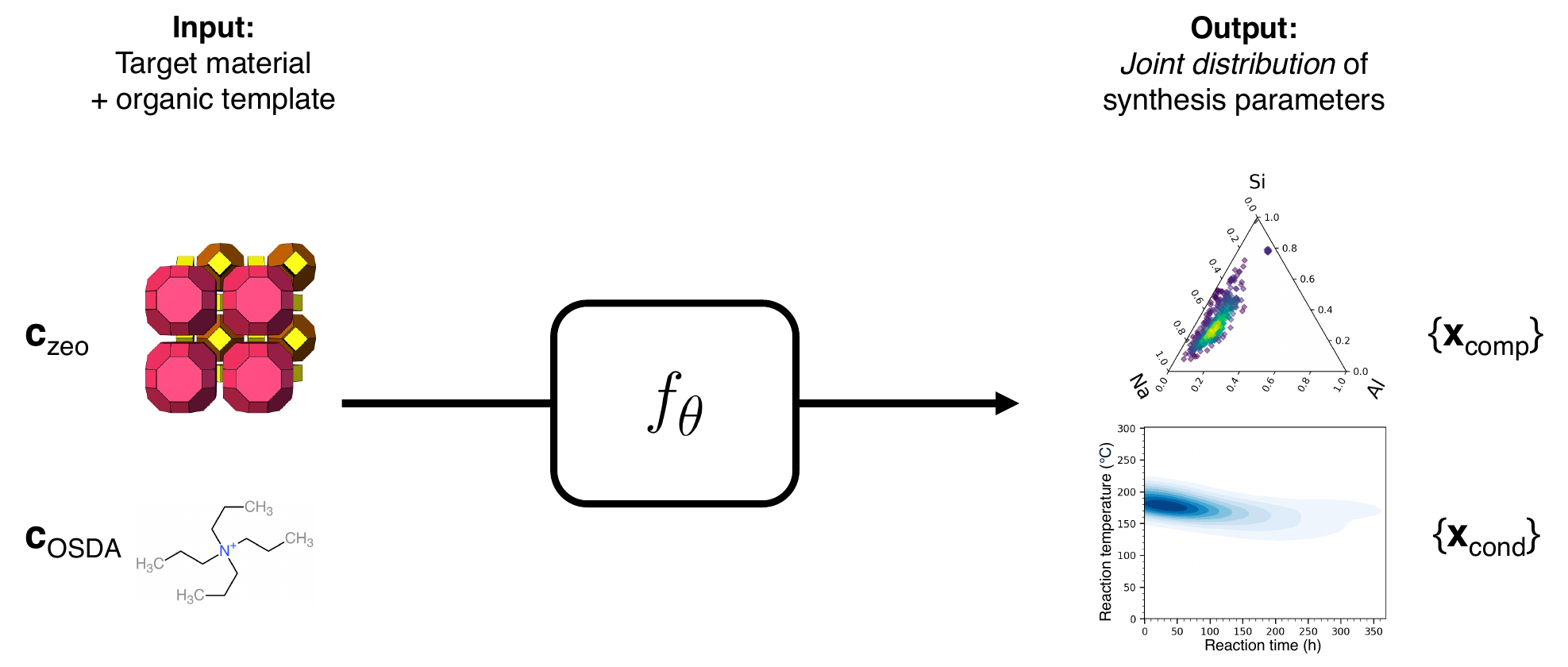}
    \caption{\textbf{Materials synthesis prediction task.} Given a target zeolite material $c_{\text{zeo}}$ and an OSDA $c_{\text{OSDA}}$, the task is to predict an \textit{ensemble} of synthesis routes consisting of gel compositions $\{x_{\text{comp}}\}$ (e.g. Si/Al) and reaction conditions $\{x_{\text{cond}}\}$ (e.g. reaction temperature/time).
    }
    \label{fig:task}
\end{figure}

\section{Coverage metric (COV-F1)}
\label{coverage_metrics}
We define 2 metrics to compare two distributions of synthesis parameters: Synthesis routes generated by a model $\left\{\boldsymbol{S}_k\right\}_{k \in[1 . . . K]}$ and synthesis routes reported in literature $\left\{\boldsymbol{S}_l\right\}_{l \in[1 . .  . L]}$. This method is inspired by \cite{xie2021crystal}.

For each synthesis parameter, we calculate the Euclidean distance between generated and literature synthesis parameter, which we call $D_{\text{syn}}(\boldsymbol{S}_k, \boldsymbol{S}_l)$. Following the well-known classification metrics of precision and recall, we define the coverage
metrics as:

\begin{equation}
    \begin{aligned}
    \text{COV-R (Recall)}=\frac{1}{L} 
    \Big|\{l \in[1 . . L]: \exists k \in[1 . . K], 
    D_{\text{syn}}(\boldsymbol{S}_k, \boldsymbol{S}_l)
    <\delta_{\text{syn}} \}\Big|
    \end{aligned}
\end{equation}

\begin{equation}
    \begin{aligned}
    \text{COV-P (Precision)}=\frac{1}{K} 
    \Big| \{k \in[1 . . K]: \exists l \in[1 . . L], 
    D_{\text{syn}}(\boldsymbol{S}_l, \boldsymbol{S}_k)
    <\delta_{\text{syn}} \} \Big|
    \end{aligned}
\end{equation}

where COV is Coverage. The recall metrics measure how many ground synthesis routes are correctly recalled by the generated synthesis routes, while the precision metrics measure how many generated synthesis routes are probable (close to literature-reported routes). The threshold $\delta_{\text{syn}}$ is defined by a domain expert in zeolite synthesis in Table \ref{thresholds}. Finally, the overall metric is COV-F1, which is defined as the harmonic mean of COV-R and COV-P i.e. $\text{COV-F1} =  \frac{2}{\frac{1}{\text{COV-R}} + \frac{1}{\text{COV-P}}}$, which ranges from $0-1$ (higher is better). An ideal generative model for materials synthesis planning should maximize its quality of generated synthesis routes (high precision) while maximizing their diversity (high recall). As such, COV-F1 is used as the metric to evaluate the models.

\section{Model implementation}
\label{model_implementation}
\textbf{Diffusion model training and sampling} We train a conditional DDPM with U-Net \cite{ronneberger2015u} as the score function, with an input dimension of 12, followed by subsequent layers dimensions multipliers of 128, 64 and 32 for downsampling and the reverse for upsampling. The score function is trained with $T = 1000$, with a $p_{\text{uncond}}= 0.1$ for classifer-free guidance with exponential moving average ($\beta = 0.99$), batch size of 32 at a constant learning rate of $4 \times 10^{-4}$ for 1M epochs. For sampling, $w = 1.0$ was found to be optimal (see Fig. \ref{fig:p_uncond_and_cond_scale}). Models are trained on a single NVIDIA RTX A5000 GPU with a wall time of 51 hours.

\textbf{Equivariant graph encoder}
We implement an equivariant graph neural network (EGNN) using the \texttt{e3nn} package (inspired by \cite{sheriff2024chemical}) to encode the structure of the zeolite as a graph. Here, we use a single layer EGNN with $l_{\text{max}} = 2$, $n_{\text{neighbors}} = 10$ and $n_{\text{nodes}} = 140$. In this case, we assume a pure-silica zeolite structure obtained from the IZA database \cite{baerlocher2008database}, and initialize node embeddings as one-hot according to the element identity (Si or O). We found that the EGNN did not outperform the invariant features (Wasserstein distance of 0.60 for EGNN vs. 0.53 for invariant features). A plausible reason for this result is as follows: The invariant geometric descriptors (e.g., pore volume, cage diameters, framework density) are physically meaningful and capture key aspects of the zeolite structure; these hand-crafted features may already provide a sufficiently informative representation for the synthesis prediction task. In contrast, the EGNN must learn these structural features from scratch. Given the limited size of the dataset (233 distinct zeolite frameworks), the EGNN-learned representation may not generalize better than the invariant descriptors. Considering that the EGNN encoder takes significantly longer time during training and inference, we use invariant features for \texttt{DiffSyn} and all baseline models.

\begin{figure}[t]
    \centering
    \includegraphics[width=\linewidth]{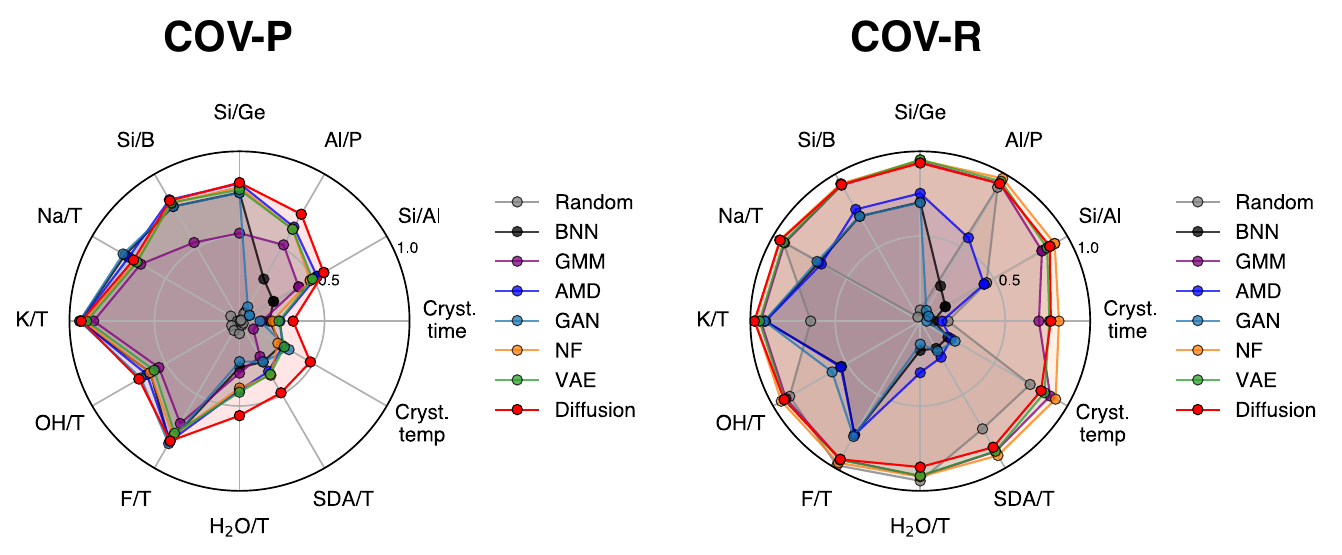}
    \caption{\textbf{COV-P (precision) and COV-R (recall) for all models.} COV-F1 is the mean of COV-P and COV-R. Notice that high COV-F1 for generative models such as diffusion and VAE (vs. AMD) largely driven by higher COV-R, while high COV-F1 for diffusion (vs. other generative models) largely driven by higher COV-P.
    }
    \label{fig:precision_recall}
\end{figure}

\begin{figure}[h!]
    \centering
    \includegraphics[width=\linewidth]{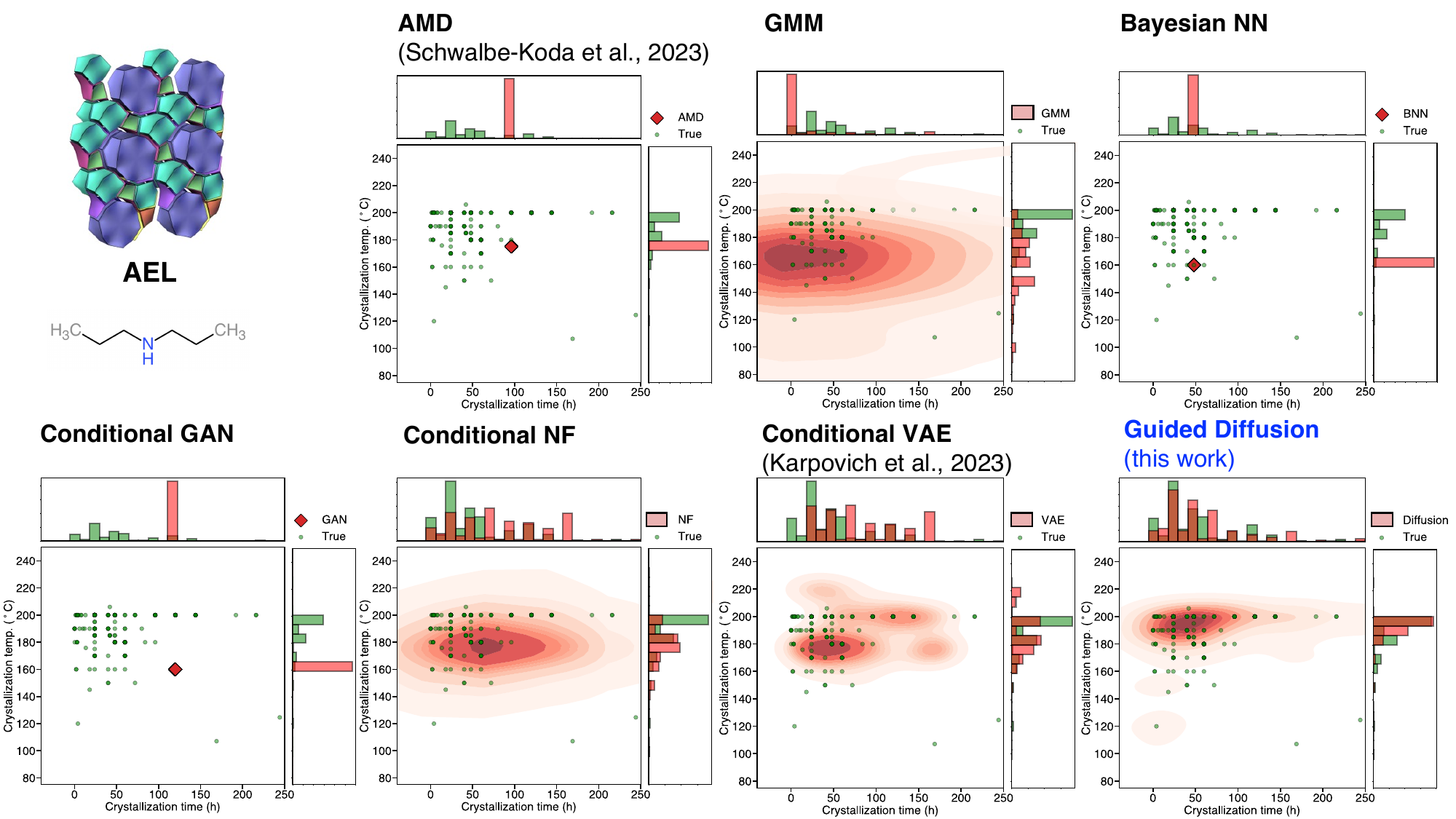}
    \caption{\textbf{Predicted $\{x_{\text{cond}}^{\text{gen}}\}$ vs. ground-truth $\{x_{\text{cond}}^{\text{true}}\}$ reaction conditions across different models} on an \textit{unseen} zeolite-OSDA system (AEL, dipropylamine). Here, the reaction conditions are crystallization temperature and time. As shown, prior methods based on regression models (AMD) and generative models (Conditional GAN, normalizing flow and VAE) are unable to accurately capture the ground-truth. In contrast, our guided diffusion method (blue, bottom right) is able to generate reaction conditions (red heatmap) that accurately captures the distribution of literature-reported conditions (green points). However, a minority (see 2 points at bottom right) are not captured by the diffusion model. This two points are outliers (very low crystallization temperature and very long crystallization time).
    }
    \label{fig:2d_dist}
\end{figure}

\begin{figure}[h!]
    \centering
    \includegraphics[width=\linewidth]{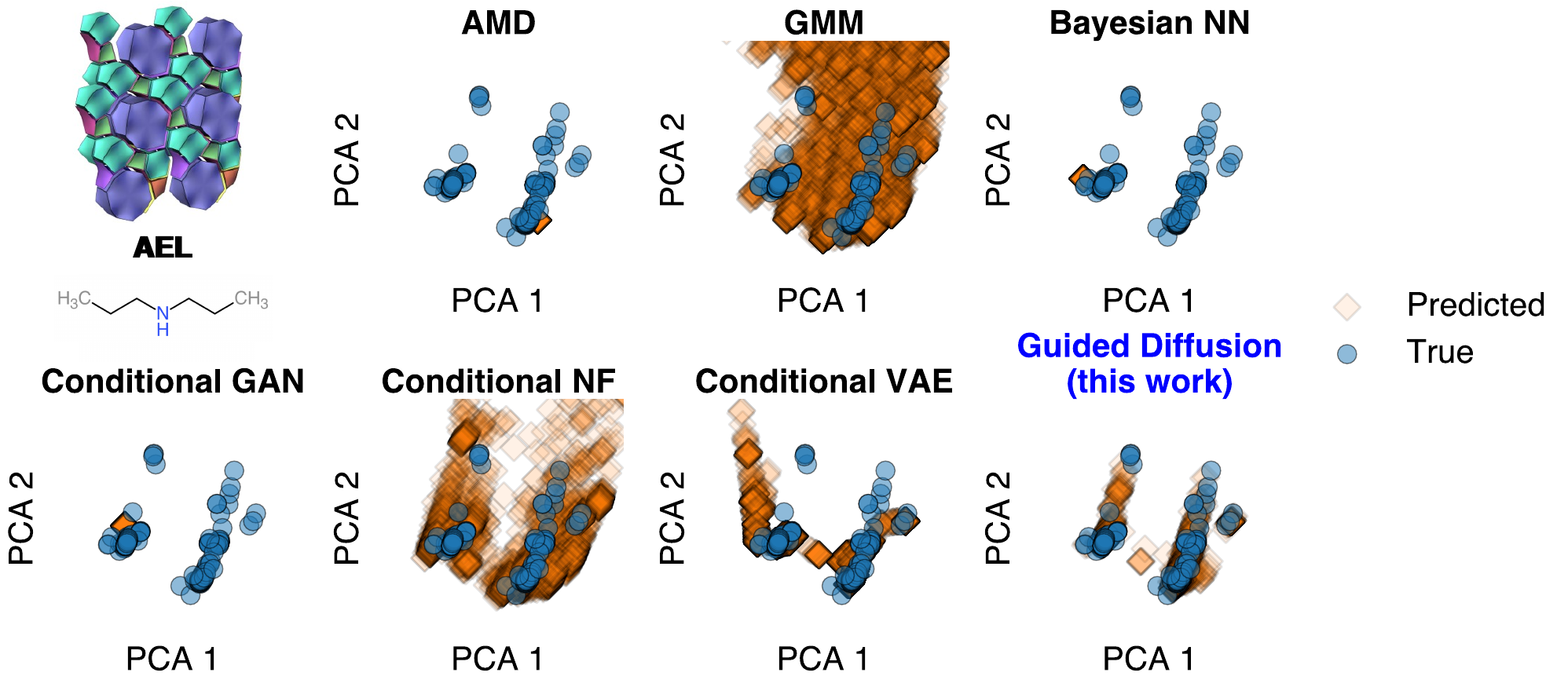}
    \caption{\textbf{Principal component analysis of predicted vs. ground-truth synthesis routes across different models} on an \textit{unseen} zeolite-OSDA system (AEL, dipropylamine). This plot demonstrates the \textit{multi-modal} nature of synthesis routes for a given structure, where the true distribution (blue) have 2 major modes. Regression-based models (AMD, Bayesian NNs) predict only one of the major modes while missing the other. GAN suffers from mode collapse to one of the modes. NF and VAE are able to predict both modes, but also generated a large number of false positives. In contrast, only guided diffusion (\texttt{DiffSyn}) is able to accurately predict the true distribution of synthesis parameters.
        }
    \label{fig:2d_pca_compare_with_struct}
\end{figure}

\begin{figure}[h!]
    \centering
    \includegraphics[width=0.8\linewidth]{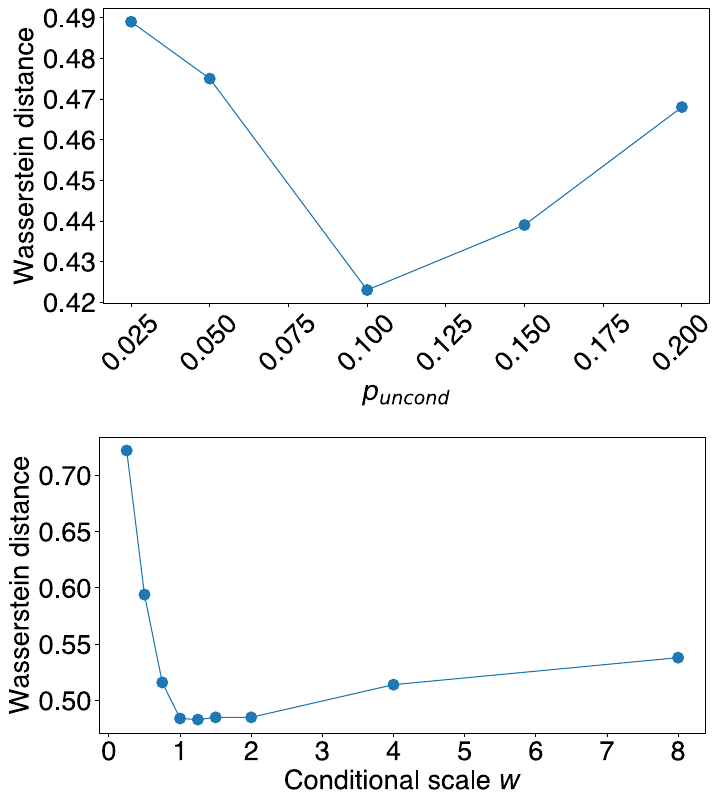}
    \caption{\textbf{Influence of classifier-free guidance hyperparameters on model performance} in terms of $p_{\text{uncond}}$ and guidance strength $w$.
    }
    \label{fig:p_uncond_and_cond_scale}
\end{figure}

\begin{figure}[h!]
    \centering
    \includegraphics[width=\linewidth]{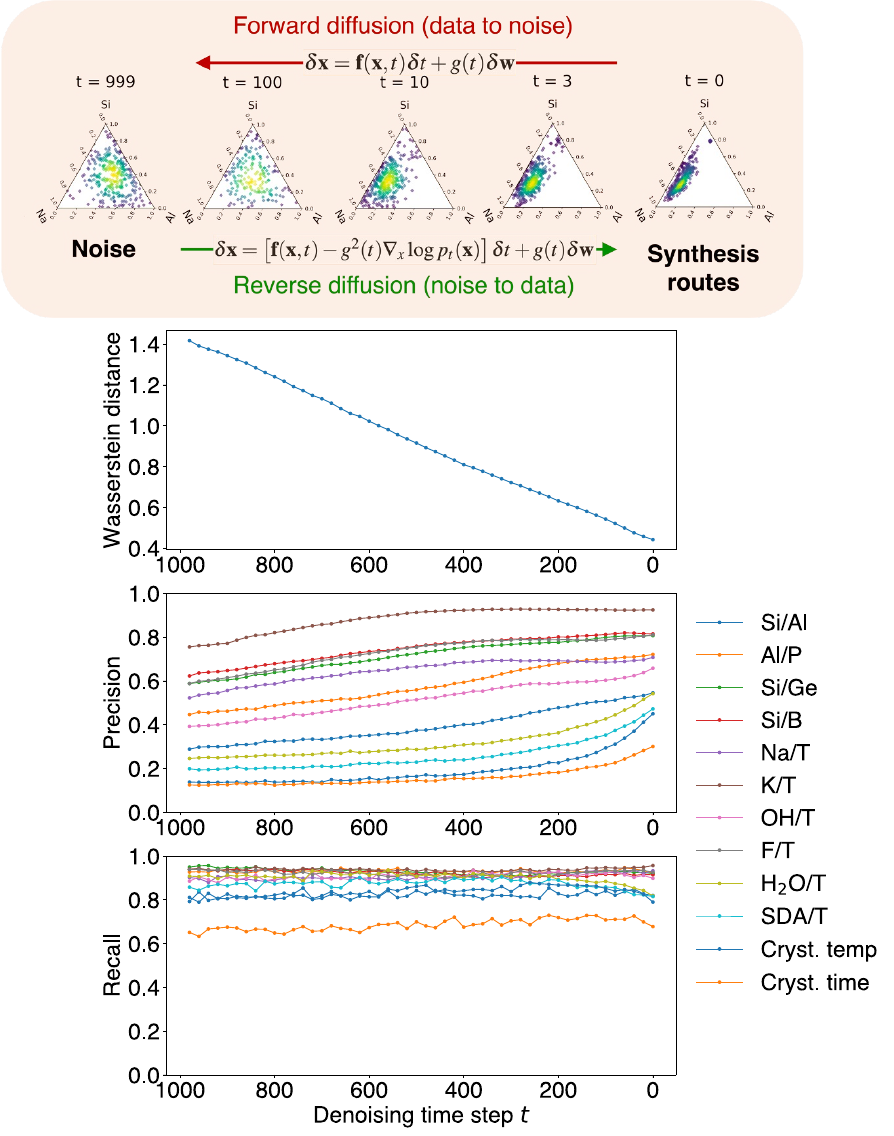}
    \caption{\textbf{Model performance across denoising diffusion trajectory} measured by Wasserstein distance ($\downarrow$), precision ($\uparrow$) and recall ($\uparrow$). At inference, reverse diffusion (green arrow) maps noise (at $t=T$) to probable synthesis routes (at $t=0$), leading to a monotonic improvement in Wasserstein distance (between generated and ground truth distributions) and precision (COV-P). In contrast, recall (COV-R) remains relatively constant across the trajectory. This shows that, in reverse time, the diffusion model is able to generate higher quality synthesis recipes while maintaining their diversity, hence capturing the \textit{one-to-many} structure-synthesis relationship for zeolite materials.
    }
    \label{fig:metrics_vs_t}
\end{figure}

\begin{figure}[h!]
    \centering
    \includegraphics[width=\linewidth]{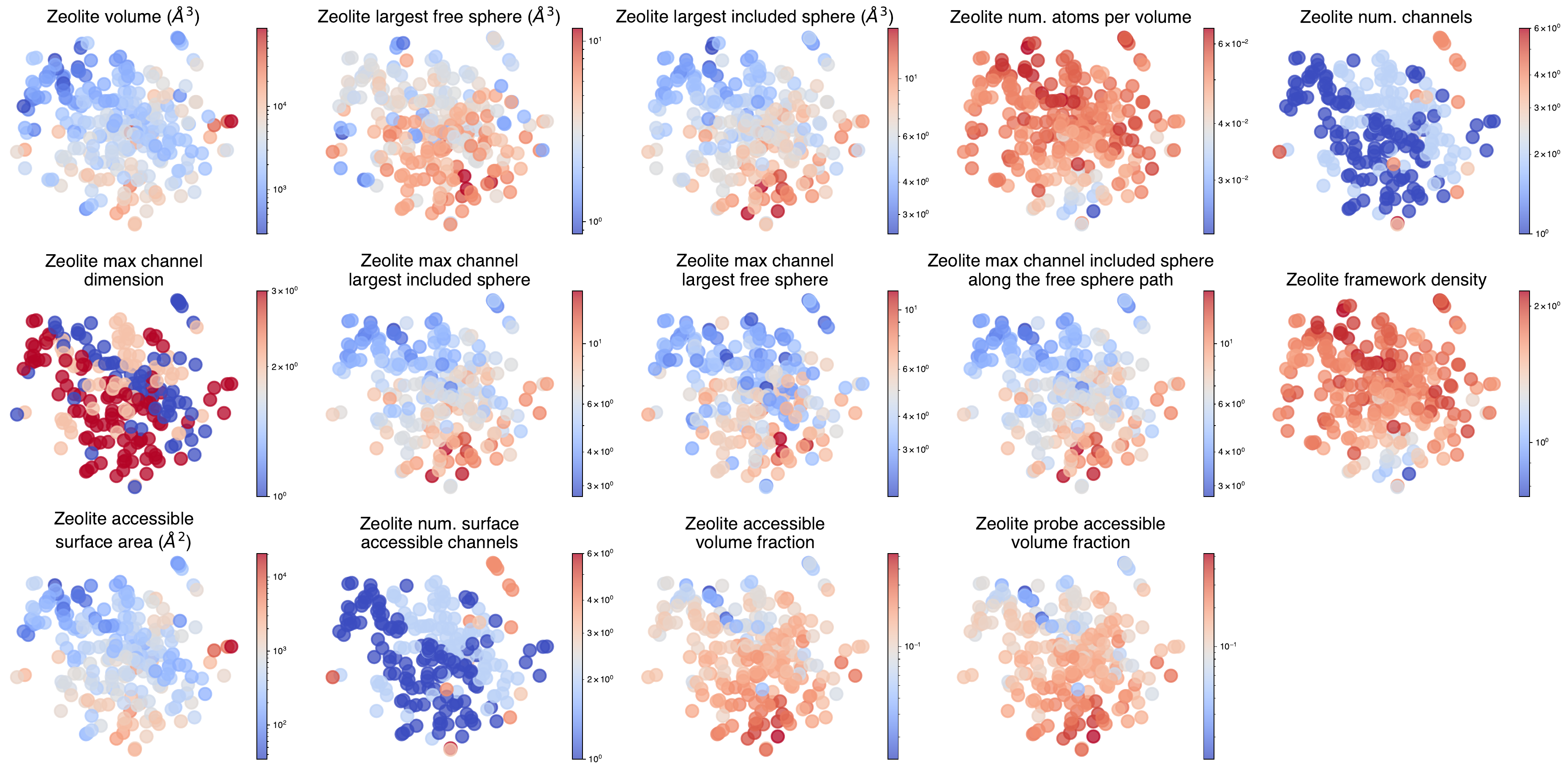}
    \caption{\textbf{Learned zeolite representations are chemically meaningful.} Latent space is smooth and continuous with respect to zeolite physical properties (e.g., accessible surface area).
    }
    \label{fig:zeolite_embeddings}
\end{figure}

\begin{figure}[h!]
    \centering
    \includegraphics[width=\linewidth]{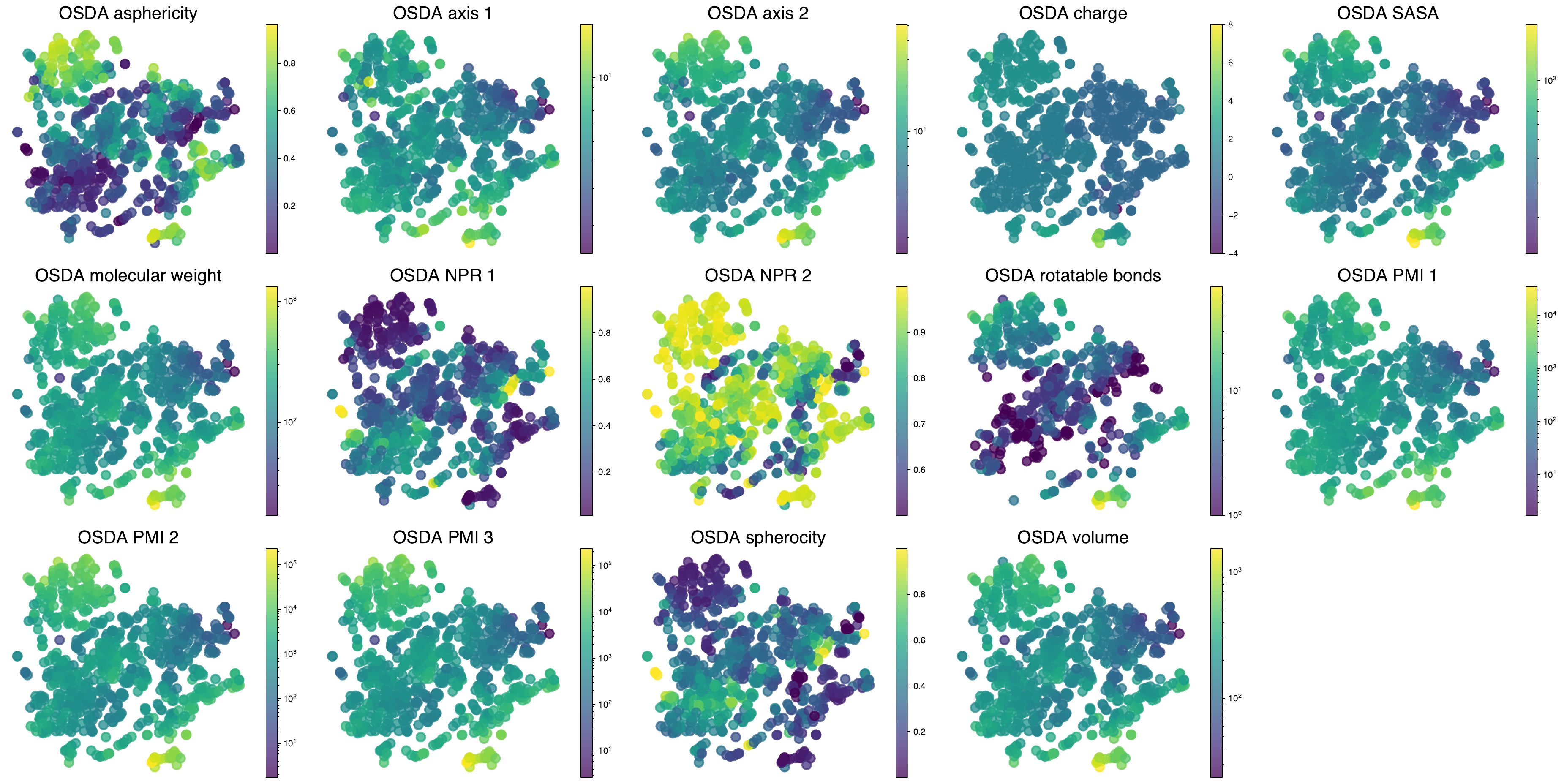}
    \caption{\textbf{Learned OSDA representations are chemically meaningful.} Latent space is smooth and continuous with respect to OSDA physical properties (e.g., volume).
    }
    \label{fig:osda_embeddings}
\end{figure}

\begin{figure}[h!]
    \centering
    \includegraphics[width=\linewidth]{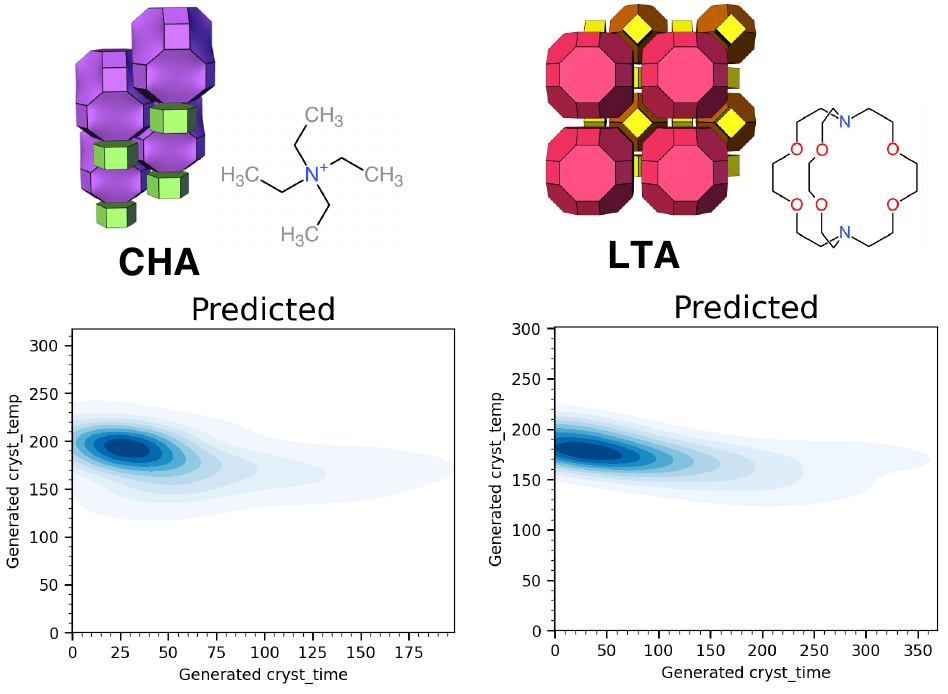}
    \caption{\textbf{Model captures physical relationships between synthesis parameters.} In the above 2 examples, generated crystallization temperatures and times follow an inverse relationship, which is aligns with the Arrhenius equation $k = Ae^{\frac{-E_a}{RT}}$, where rate constant $k$ (determines crystallization time) is inversely related to synthesis temperature. $A$, $E_a$, $R$ and $T$ refer to the pre-exponential factor, activation energy, gas constant and temperature, respectively.
    }
    \label{fig:physical_relationship}
\end{figure}

\begin{figure}[h!]
    \centering
    \includegraphics[width=0.63\linewidth]{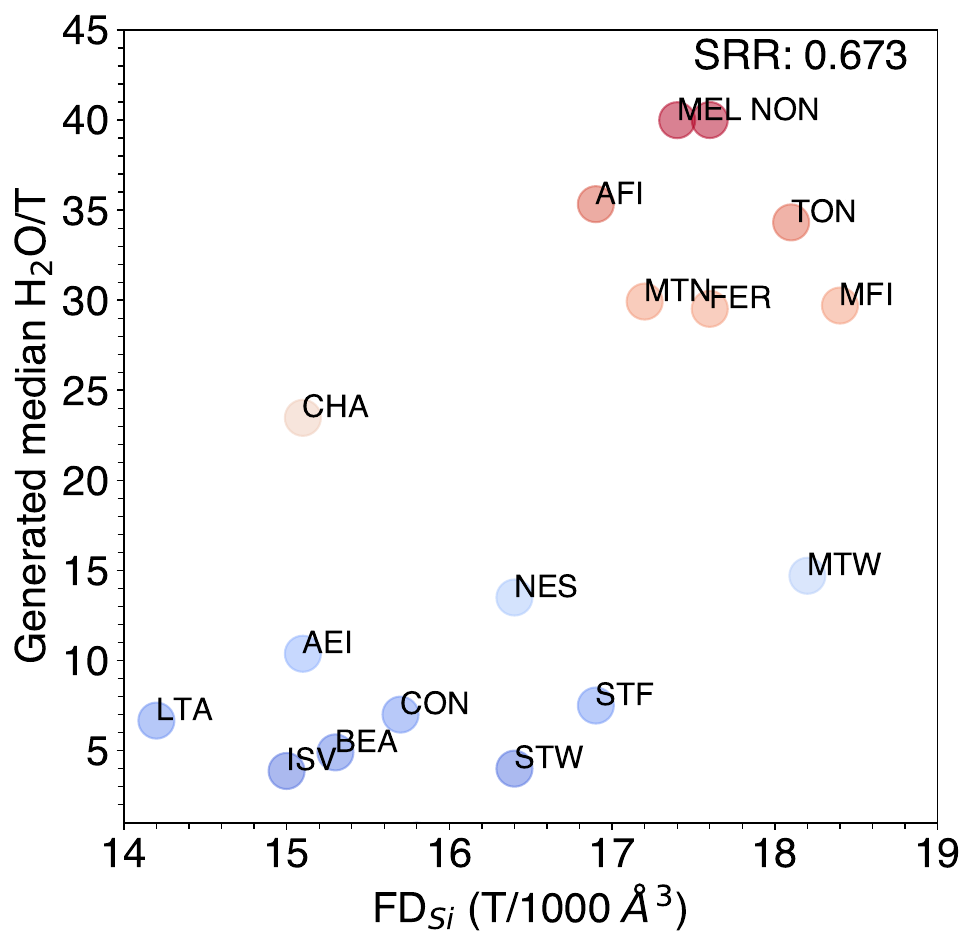}
    \caption{\textbf{Model captures domain-specific heuristics.} A positive correlation (Spearman's rank coefficient: 0.673) exists between median diffusion-generated H$_2$O/T and framework density (FD$_{\text{Si}}$) of zeolite structure for fluoride-mediated synthesis of high silica (Si/Al $>$ 30) aluminosilicates. This agrees with Villaescusa's rule \cite{camblor1999synthesis}, which states that denser phases (higher FD$_{\text{Si}}$) are favored at the less concentrated conditions (higher H$_2$O/T), showing that the model has learned domain-specific rules in zeolite synthesis.
    }
    \label{fig:h2o_vs_fwd}
\end{figure}

\begin{figure}[h!]
    \centering
    \includegraphics[width=0.70\linewidth]{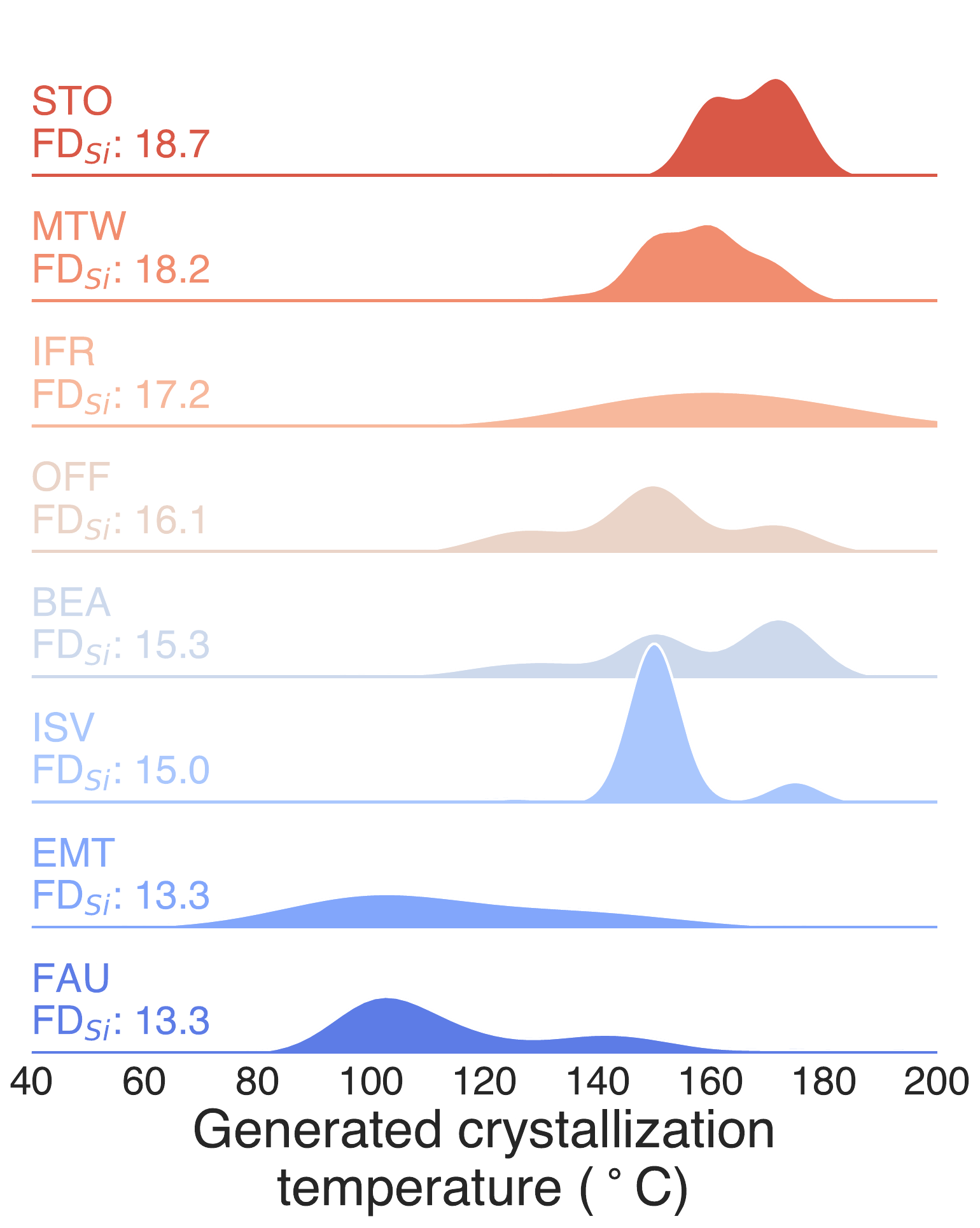}
    \caption{\textbf{Model predictions follow thermodynamics of zeolite formation.} A positive correlation (Spearman's rank coefficient: 0.931) exists between median diffusion-generated crystallization temperature and framework density (FD$_{\text{Si}}$) of zeolite structure for large-pore aluminosilicates. This can be seen in an rightward shift in distribution of generated temperatures as FD$_{\text{Si}}$ increases. This agrees with the thermodynamic argument that higher crystallization temperatures enable the synthesis to overcome the energy activation barrier to form more stable structures with higher framework density \cite{le2019process, pan2024zeosyn}. Furthermore, this observation aligns with Ostwald’s rule of stages, where the zeolite passes through metastable states before reaching the most thermodynamically favorable framework.
    }
    \label{fig:temp_vs_fwd}
\end{figure}

\begin{figure}[h!]
    \centering
    \includegraphics[width=\linewidth]{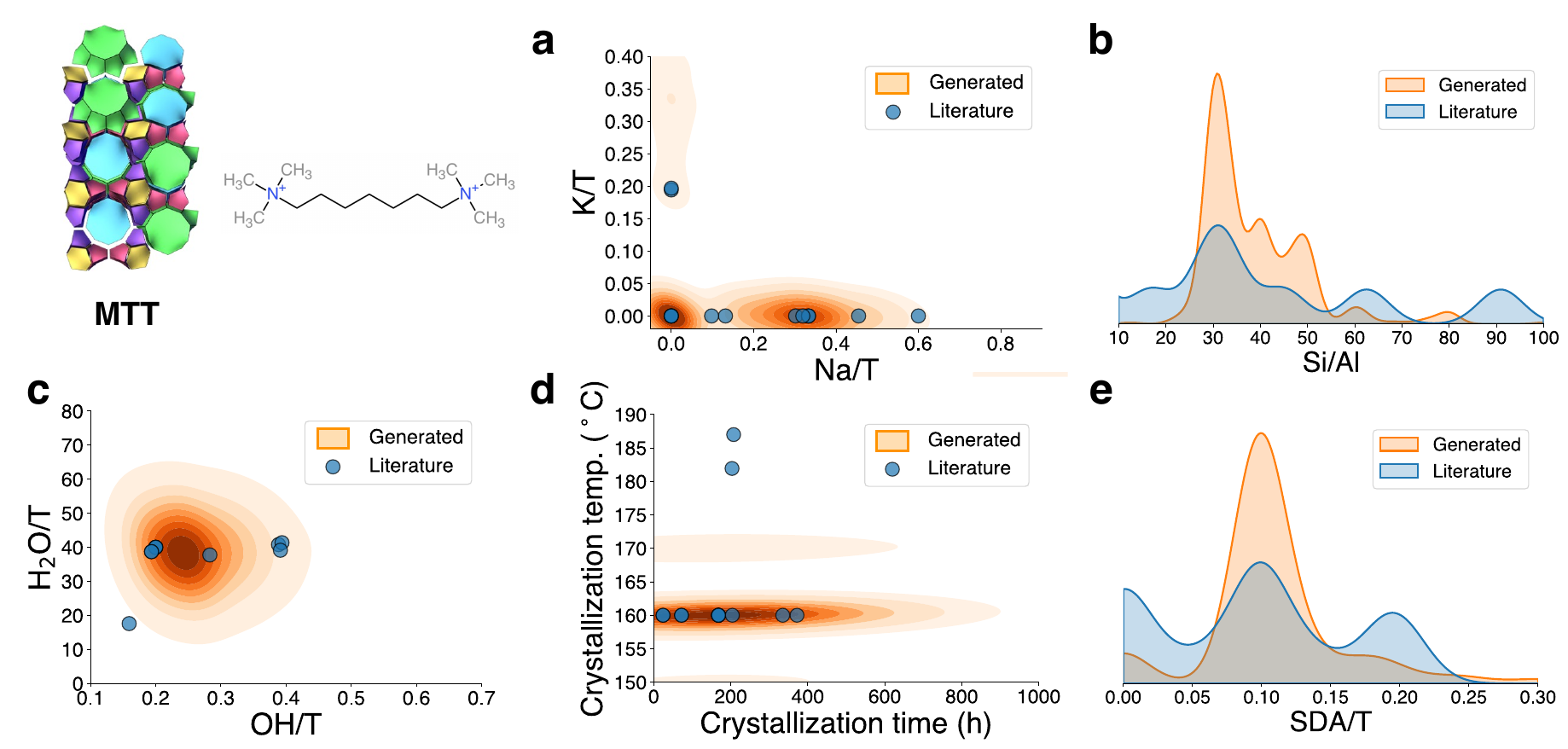}
    \caption{\textbf{Case study: MTT synthesis.} Generated synthesis parameters for MTT (in cluster 6 of Fig. \ref{fig:fig3}a) overlap closely with unseen synthesis parameters reported in literature. While the model is accurate for most parameters, it is unable to capture very low and high values of Si/Al for MTT (see \textbf{(b)}).
    }
    \label{fig:CS_MTT}
\end{figure}

\begin{figure}[h!]
    \centering
    \includegraphics[width=\linewidth]{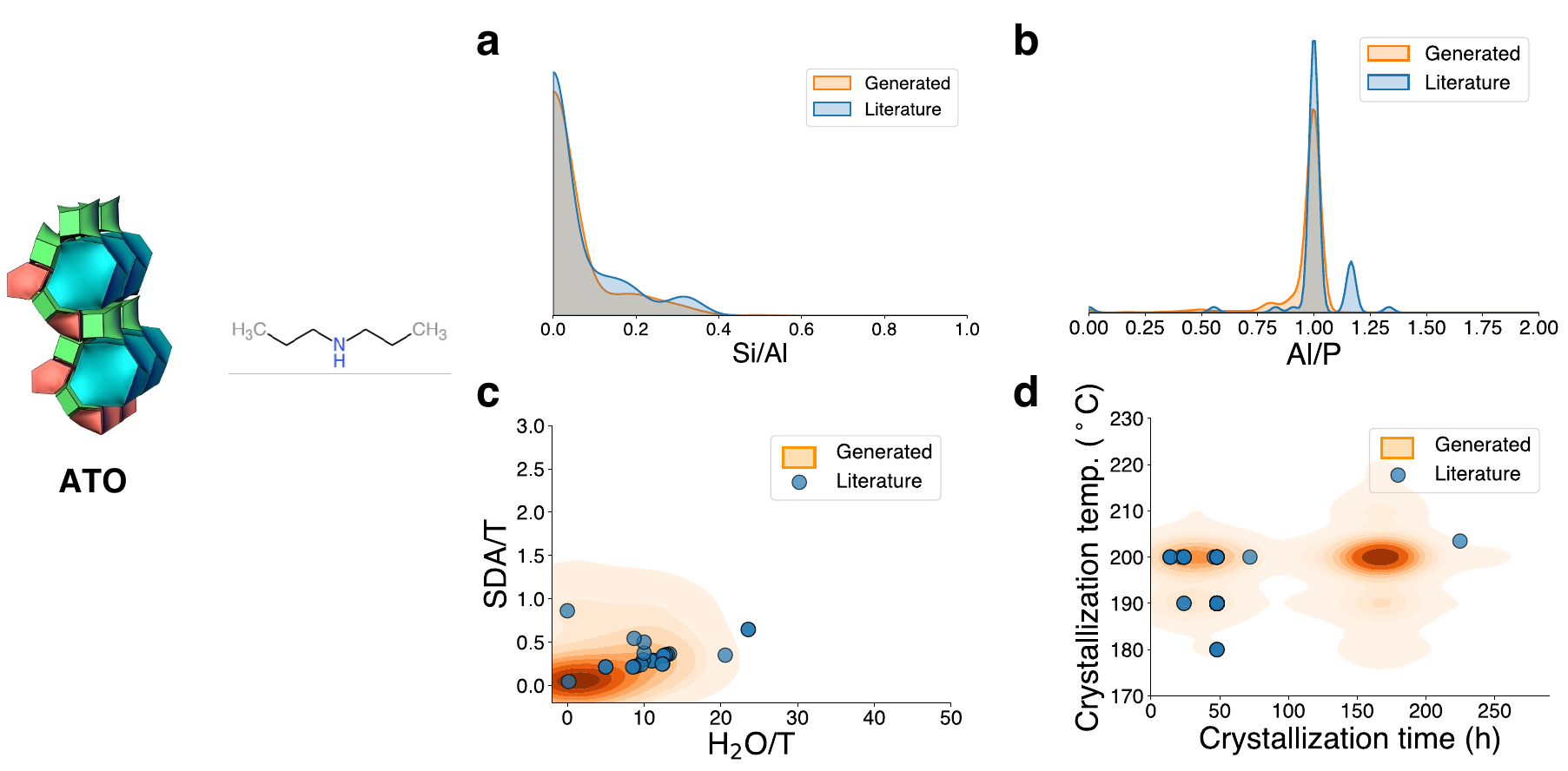}
    \caption{\textbf{Case study: ATO synthesis.} Generated synthesis parameters for ATO (in cluster 1 of Fig. \ref{fig:fig3}a) overlap closely with unseen synthesis parameters reported in literature. While the model is accurate for most parameters, it predicts an extra peak at crystallization time = 170h and crystallization temperature = 200$^\circ C$ that has weak overlap with only 1 literature datapoint (see \textbf{(d)}). This demonstrates a potential drawback of using literature reported synthesis routes as validation, as they are partial ground-truths i.e. A specific synthesis route that is not being observed in literature is \textit{not a sufficient} condition that the route will fail. Consequently, this will increase the false positive rate, and reduce the COV-R metric (see Fig. \ref{fig:precision_recall}) of our model. As such, the reported values of the COV-R metric in Fig. \ref{fig:precision_recall} are lower bounds. 
    }
    \label{fig:CS_ATO}
\end{figure}

\begin{figure}[h!]
    \centering
    \includegraphics[width=\linewidth]{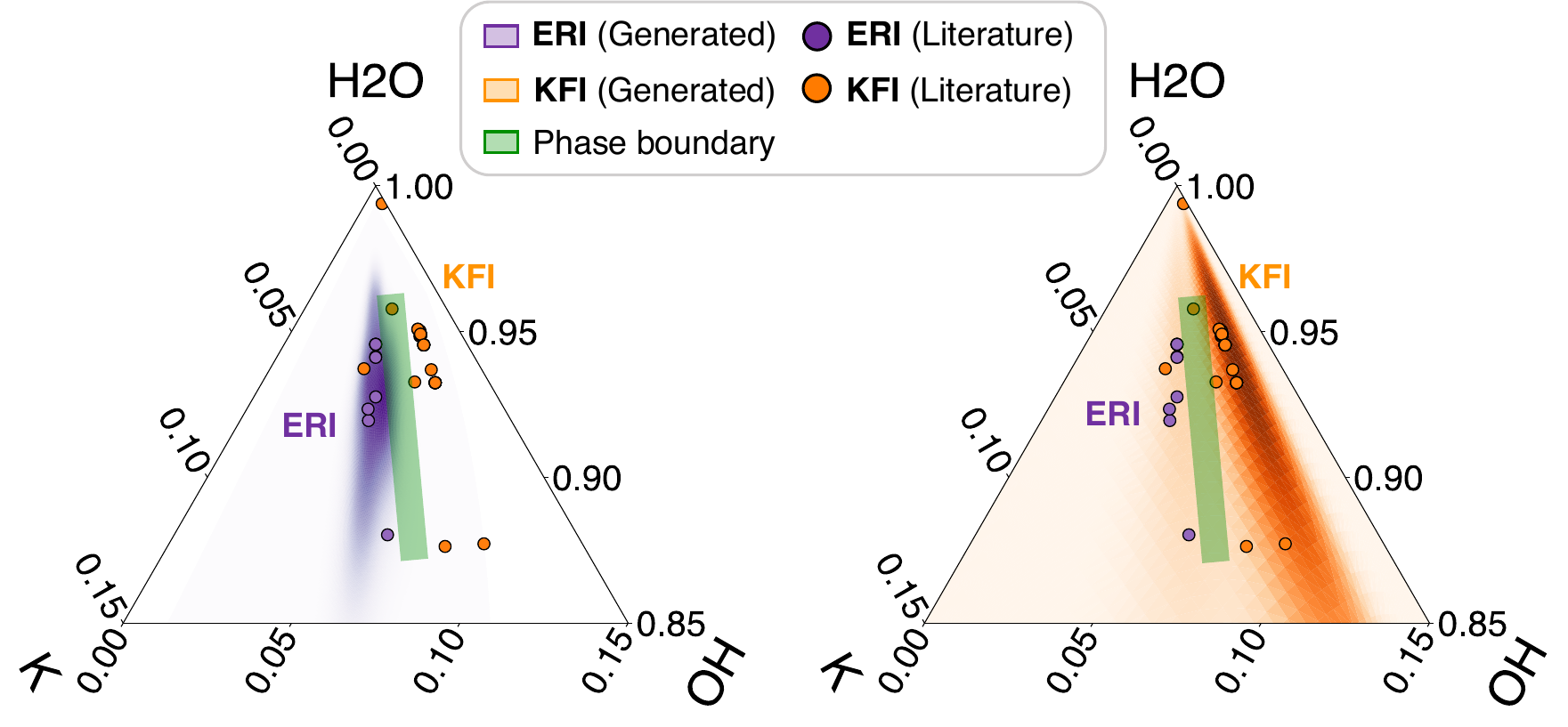}
    \caption{\textbf{Competing phases ERI and KFI.} Heatmaps refer to generated routes, while points refer to literature-reported synthesis routes. Notice that the model accurately predicts the phase boundary (green shaded region) between ERI and KFI.
    }
    \label{fig:ERI_KFI}
\end{figure}

\begin{figure}[h!]
    \centering
    \includegraphics[width=\linewidth]{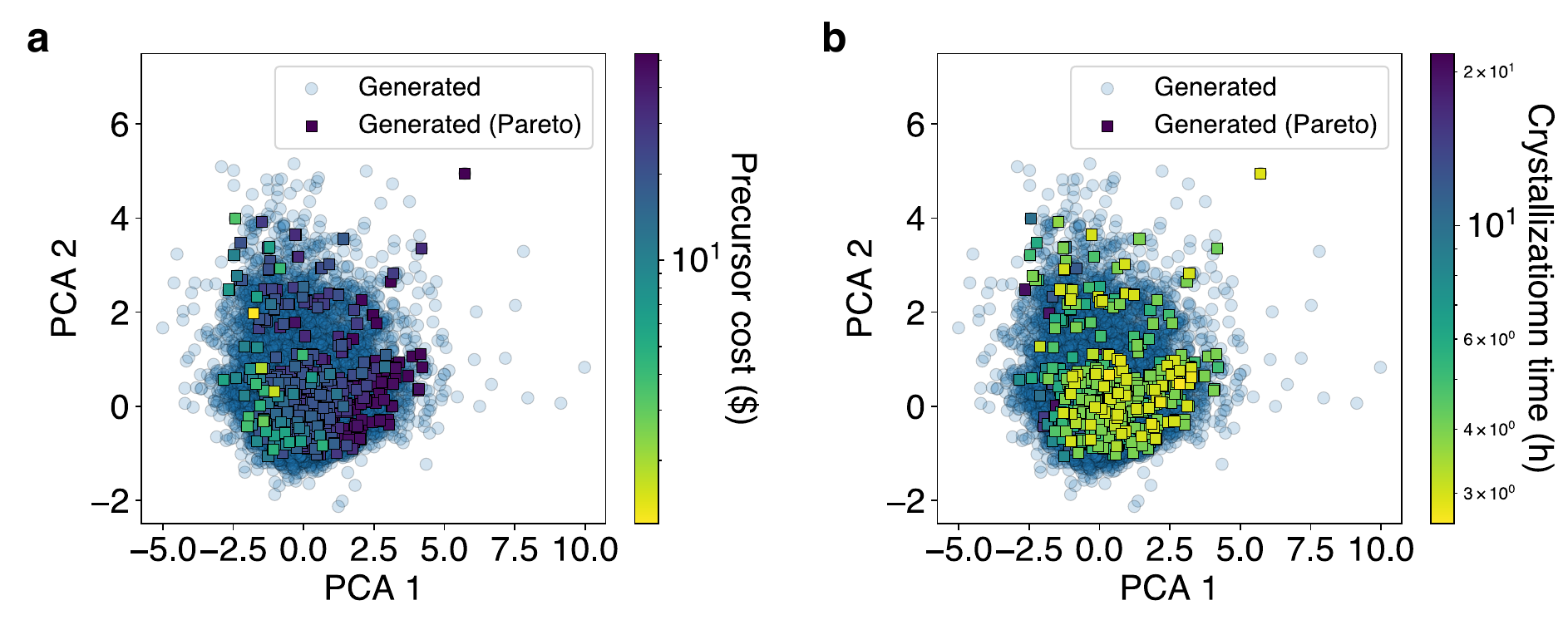}
    \caption{\textbf{PCA of synthesis space for CHA-TMAda.} Circles refer to all generated synthesis routes. Squares refer to generated synthesis routes at/near the Pareto front in discussed in Fig. \ref{fig:fig4}b and Section \ref{sec:optimal}. Each point on the Pareto front is colored by \textbf{(a)} precursor cost \textbf{(b)} crystallization time. Notice that majority of Pareto-optimal synthesis routes reside only in a bottom-half region, and the each of the two objectives (precursor cost and crystallization time) are optimal on \textit{opposite} sides (left for low precursor cost, right for low crystallization time) of the synthesis space.
    }
    \label{fig:pca_cost_time}
\end{figure}

\begin{figure}[h!]
    \centering
    \includegraphics[width=\linewidth]{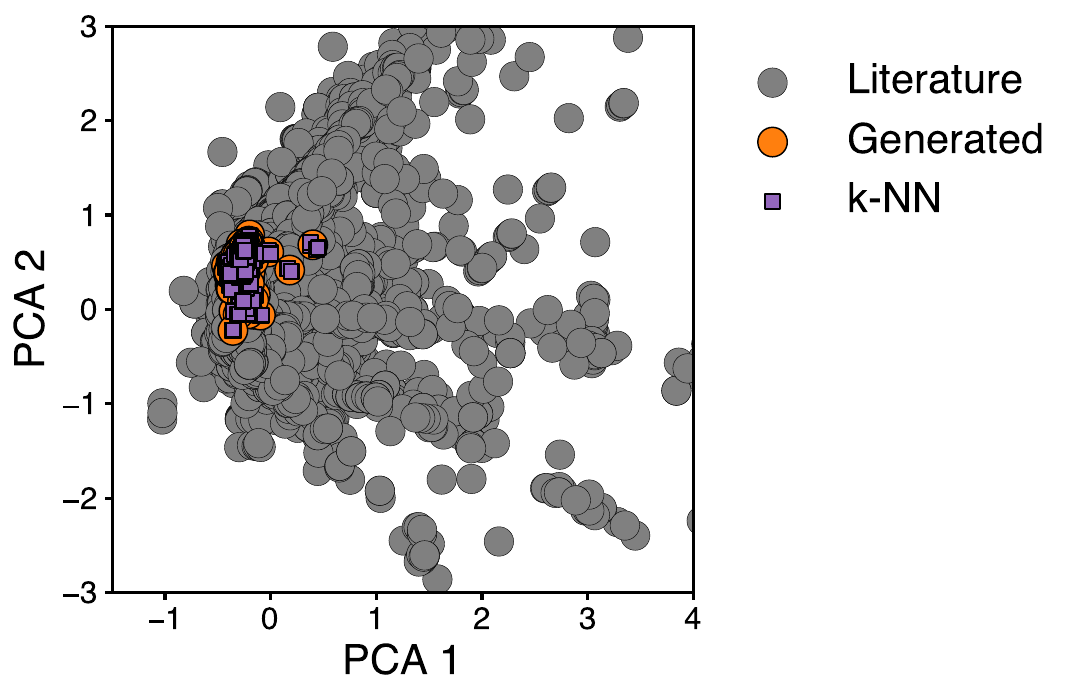}
    \caption{\textbf{PCA of synthesis parameters} of all routes reported in literature (grey), \texttt{DiffSyn}-generated for UFI (orange) and $k$-nearest neighbors ($k=5$) of generated UFI synthesis in literature-reported synthesis.
    }
    \label{fig:pca_knn}
\end{figure}

\begin{figure}[h!]
    \centering
    \includegraphics[width=0.7\linewidth]{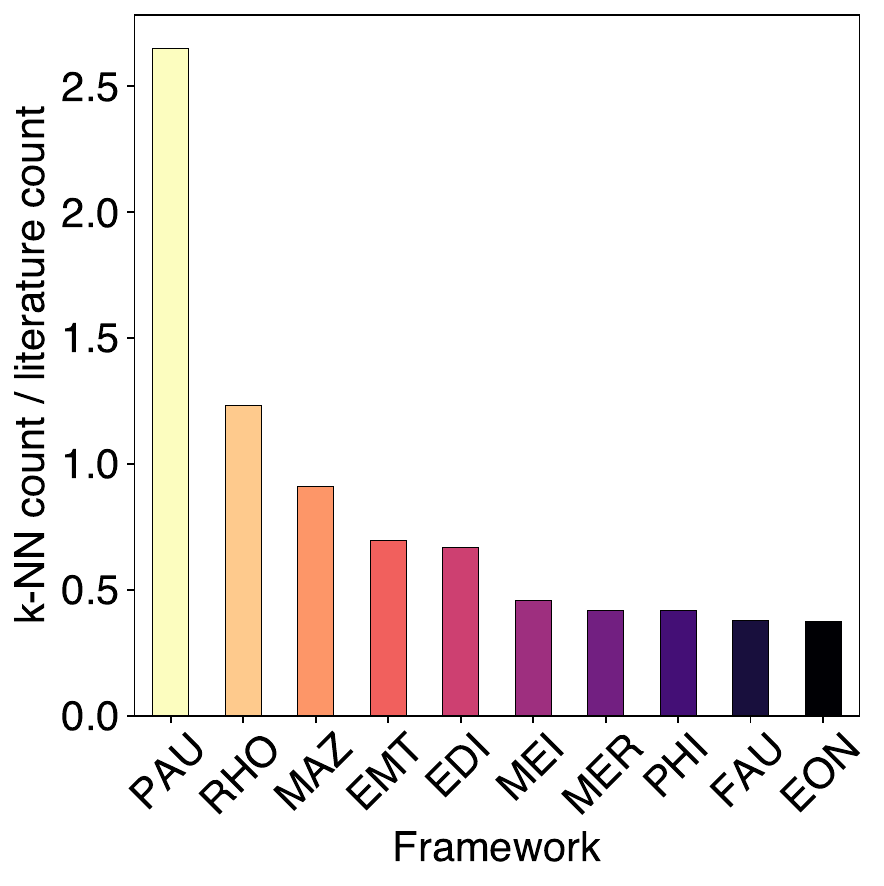}
    \caption{\textbf{Normalized frequency of $k$-NN recipes} (purple squares in Fig. \ref{fig:pca_knn}). Note, it is important to normalize the k-NN count via dividing by the framework’s frequency in literature. Larger values ($k$-NN count / literature count) suggest higher ‘synthesis similarity’ to the \texttt{DiffSyn}-generated UFI recipes. Notably, the top 2 most similar frameworks (PAU and RHO) share a common composite building unit (\textit{lta}) with UFI. Beyond this, we did not observe a clear trend in terms of common CBUs for the remaining frameworks. This may suggest that there is not a clear correlation between structure and synthesis.
    }
    \label{fig:knn_zeo_freq}
\end{figure}

\begin{figure}[h!]
    \centering
    \includegraphics[width=0.65\linewidth]{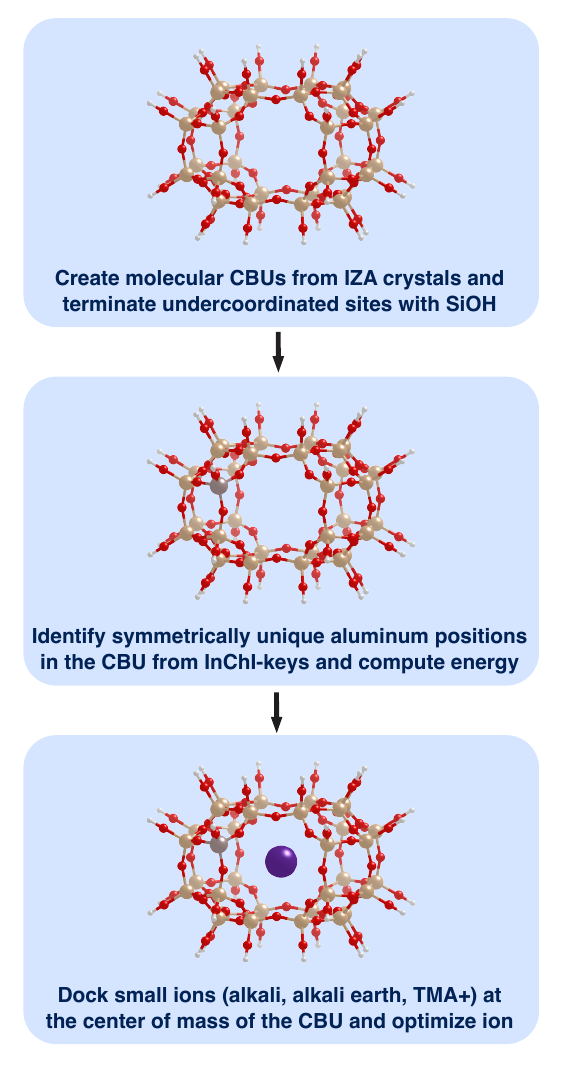}
    \caption{\textbf{DFT workflow for docking ions in zeolite composite building units (CBUs).} First, CBUs are extracted from zeolite structures (from IZA database \cite{baerlocher2008database}) with SiOH capping undercoordinated sites. Second, an exhaustive enumeration of unique aluminum occupancies in the CBU is performed, followed by DFT energy calculations. Third, ions are docked in the CBU to calculate the binding energy between ions and CBU.
    }
    \label{fig:dft_docking}
\end{figure}

\begin{table}[h!]
  \caption{\textbf{Expert-defined thresholds $\delta_{\text{syn}}$.} Values are based on practical utilty and experimental error in zeolite synthesis.}
  \centering
  \begin{tabular}{p{0.15\linewidth}  p{0.2\linewidth}}
    \toprule
    \textbf{Synthesis parameter} & \textbf{$\delta_{\text{syn}}$ (unitless unless otherwise specified)} \\
    \midrule
    Si/Al & 1.5 \\
    \midrule
    Al/P & 0.05 \\
    \midrule
    Si/Ge & 0.05 \\
    \midrule
    Si/B & 0.05 \\
    \midrule
    Na$^+$/T & 0.05 \\
    \midrule
    K$^+$/T & 0.05 \\
    \midrule
    OH$^-$/T & 0.05 \\
    \midrule
    F$^-$/T & 0.05 \\
    \midrule
    H$_{\text{2}}$O/T & 5 \\
    \midrule
    SDA/T & 0.05 \\
    \midrule
    Crystallization temperature & 5 $^\circ$C \\
    \midrule
    Crystallization time & 12 h \\
    
    \bottomrule
  \end{tabular}
  \label{thresholds}
\end{table}

\begin{table}[h!]
  \caption{\textbf{Cost of precursors used in synthesis.} Precursor cost $=\sum_i \eta_i$ (Source: \cite{muraoka2020multi})}
  \centering
  \begin{tabular}{p{0.35\linewidth}  p{0.2\linewidth}}
    \toprule
    \textbf{Reagent} & \textbf{Cost parameter $\eta$} \\
    \midrule
    Colloidal silica (LUDOX AS-40) & 0.036 \\
    \midrule
    Aluminum hydroxide & 0.052 \\
    \midrule
    NaOH & 0.0087 \\
    \midrule
    NaCl & 0.0069 \\
    \midrule
    OSDA & 190 \\
    \bottomrule
  \end{tabular}
  \label{precursor_cost}
\end{table}

\begin{table}[h!]
  \caption{\textbf{Physicochemical descriptors of OSDAs.} (Source: \cite{pan2024zeosyn}).}
  \centering
  \begin{tabular}{p{0.2\linewidth}  p{0.6\linewidth}}
    \toprule
    \textbf{OSDA descriptor} & \textbf{Description} \\
    \midrule
    Asphericity & An anisometry descriptor for the deviation from the spherical shape \\
    \midrule
    Axis 1 & Two-dimensional (2D) shape descriptors of molecule calculated by projecting the atomic coordinates into a 2D space based on a principal component analysis (PCA) of the positions. The range of the distribution of points in each principal component is reported as the axis of the conformer. Axis 1 is reported as the larger axis, whereas Axis 2 is the smaller axis \\
    \midrule
    Axis 2 & See above \\
    \midrule
    Charge & Formal charge of molecule \\
    \midrule
    SASA & Solvent-accessible surface area (SASA) is the surface area of a molecule that is accessible to a solvent \\
    \midrule
    Molecular weight & Molecular mass of molecule \\
    \midrule
    NPR 1 & Normalized principal moments ratio ($\frac{I_1}{I_3}$) where I is principal moment of inertia \\
    \midrule
    NPR 2 & Normalized principal moments ratio ($\frac{I_2}{I_3}$) where I is principal moment of inertia \\
    \midrule
    Rotatable bonds & Number of rotatable bonds in the molecule. A measure of molecular flexibility. \\
    \midrule
    PMI 1 & Principal moments of inertia (PMI) are physical quantities related to the rotational dynamics of a molecule 
    \begin{equation}
    \mathrm{I}=\sum_{i=1}^{\mathrm{A}} m_i \cdot r_i^2
    \end{equation} 
    where $A$ is the number of atoms, and $m_i$ is the atomic mass and $r_i$ is the perpendicular distance from the chosen axis of the $i$th atom of the molecule
    \\
    \midrule
    PMI 2 & See above \\
    \midrule
    PMI 3 & See above \\
    \midrule
    Sphericity & Sphericity index of molecule. A measure of how closely the shape of an object resembles that of a perfect sphere \\
    \midrule
    Volume & Molecular volume calculated by using a grid-encoding of the molecular shape using a grid spacing of 0.2 Å and 2.0 Å of margin for the boxes \\
    \bottomrule
  \end{tabular}
  \label{osda_descriptors}
\end{table}

\begin{table}[h!]
  \caption{\textbf{VAE hyperparameters.} Here, we train a $\beta$-VAE \cite{higgins2017beta}.}
  \centering
  \begin{tabular}{p{0.2\linewidth}  p{0.25\linewidth}}
    \toprule
    \textbf{Hyperparameter} & \textbf{Value} \\
    \midrule
    epochs & 5000 \\
    \midrule
    batch size & 2048 \\
    \midrule
    learning rate & 0.0001 \\
    \midrule
    $\beta$ & 0.01 \\
    \midrule
    latent dim & 2 \\
    \midrule
    encoder dims & [4096, 2048, 1024, 512] \\
    \midrule
    decoder dims & [512, 1024, 2048, 4096] \\
    \bottomrule
  \end{tabular}
  \label{cvar_hp}
\end{table}

\begin{table}[h!]
  \caption{\textbf{NF hyperparameters.} Here, we train a normalizing flow with real-valued non-volume preserving (RealNVP) transformations \cite{dinh2016density}.}
  \centering
  \begin{tabular}{p{0.2\linewidth}  p{0.25\linewidth}}
    \toprule
    \textbf{Hyperparameter} & \textbf{Value} \\
    \midrule
    epochs & 3000 \\
    \midrule
    batch size & 2048 \\
    \midrule
    learning rate & 0.0001 \\
    \midrule
    num. flows & 32 \\
    \midrule
    hidden dim & 1024 \\
    \bottomrule
  \end{tabular}
  \label{cvar_hp}
\end{table}

\begin{table}[h!]
  \caption{\textbf{GAN hyperparameters.}}
  \centering
  \begin{tabular}{p{0.2\linewidth}  p{0.25\linewidth}}
    \toprule
    \textbf{Hyperparameter} & \textbf{Value} \\
    \midrule
    epochs & 400 \\
    \midrule
    batch size & 2048 \\
    \midrule
    learning rate & 0.0001 \\
    \midrule
    latent dim & 2 \\
    \midrule
    generator dims & [64, 128, 256] \\
    \midrule
    discriminator dims & [256, 128, 64] \\
    \bottomrule
  \end{tabular}
  \label{cvar_hp}
\end{table}

\begin{table}[h!]
  \caption{\textbf{BNN hyperparameters.}}
  \centering
  \begin{tabular}{p{0.2\linewidth}  p{0.25\linewidth}}
    \toprule
    \textbf{Hyperparameter} & \textbf{Value} \\
    \midrule
    epochs & 10000 \\
    \midrule
    batch size & 8192 \\
    \midrule
    learning rate & 0.0001 \\
    \midrule
    prior $\mu$ & 0 \\
    \midrule
    prior $\sigma$ & 0.1 \\
    \bottomrule
  \end{tabular}
  \label{cvar_hp}
\end{table}

\begin{table}[h!]
  \caption{\textbf{AMD hyperparameters.} Here, we train a multi-task MLP with zeolite representation from Schwalbe-Koda et al. \cite{schwalbe2023inorganic}.}
  \centering
  \begin{tabular}{p{0.2\linewidth}  p{0.25\linewidth}}
    \toprule
    \textbf{Hyperparameter} & \textbf{Value} \\
    \midrule
    epochs & 5000 \\
    \midrule
    batch size & 4096 \\
    \midrule
    learning rate & 0.0001 \\
    \midrule
    hidden dims & [512, 128] \\
    \bottomrule
  \end{tabular}
  \label{cvar_hp}
\end{table}

\begin{table}[h!]
  \caption{\textbf{GMM hyperparameters.} Here, we use a conditional density estimator via least-squares density ratio estimation. \cite{sugiyama2010conditional}. This can be implemented using the \texttt{LSConditionalDensityEstimation} function in the \texttt{cde} package.
  }
  \centering
  \begin{tabular}{p{0.2\linewidth}  p{0.25\linewidth}}
    \toprule
    \textbf{Hyperparameter} & \textbf{Value} \\
    \midrule
    center sampling method & k-means \\
    \midrule
    num. centers &5 50 \\
    \midrule
    bandwidth & 0.5 \\
    \midrule
    regularization & 1.0 \\
    \bottomrule
  \end{tabular}
  \label{cvar_hp}
\end{table}

\begin{sidewaystable}[h!]
\centering
\begin{tabular}{|c|c|c|c|c|c|c|c|c|c|c|c|c|c|c|c|}
\hline
\multicolumn{1}{|c|}{\makecell{\textbf{Sample}\\ \textbf{info}}} & \multicolumn{10}{c|}{\makecell{\textbf{Gel}\\ \textbf{composition}}} & \multicolumn{2}{c|}{\makecell{\textbf{Synthesis} \\ \textbf{conditions}}} & \textbf{Rotation} & \textbf{Seed} & \textbf{Si/Al$_{\text{ICP}}$} \\ \hline
 & Si & Al & Ge & B & Na\textsuperscript{+} & K\textsuperscript{+} & H\textsubscript{2}O & F\textsuperscript{-} & \makecell{OSDA1\\ (K222)} & \makecell{OSDA2\\ (TMA)} & \makecell{Cryst.\\ temp. ($^\circ \text{C}$)$^*$} & \makecell{Cryst.\\ time (h)} & & & \\ \hline
UFI-1 & 1 & 0.1 & 0.0 & 0.0 & 0.2 & 0.0 & 15 & 0.0 & 0.15 & 0.05  & 175 & 168 & Yes & Yes & 14.0 \\ \hline
UFI-2 & 1 & 0.07 & 0.0 & 0.0 & 0.2 & 0.0 & 15 & 0.0 & 0.15 & 0.05 & 175 & 168 & Yes & Yes & 19.0 \\ \hline
UFI-3 & 1 & 0.1 & 0.0 & 0.0 & 0.3 & 0.0 & 15 & 0.0 & 0.35 & 0.05 & 175 & 168 & Yes & Yes & 13.6 \\ \hline
UFI-4 & 1 & 0.1 & 0.0 & 0.0 & 0.2 & 0.0 & 15 & 0.0 & 0.1 & 0.05 & 175 & 168 & Yes & Yes & 13.0 \\ \hline
\end{tabular}
\caption{\textbf{Synthesis routes of UFI zeolite.} K222: Kryptofix 222 (4,7,13,16,21,24-Hexaoxa-1,10-diazabicyclo[8.8.8]hexacosane), TMA: tetramethylammonium. $^*$ Minor mode predicted by our model at 175$^\circ \text{C}$ }
\label{table:ufi_synthesis}
\end{sidewaystable}

\end{appendices}

\end{document}